\documentclass[onecolumn,showpacs,preprintnumbers,amsmath,amssymb]{revtex4}
\usepackage{graphicx}% Include figure files
\usepackage{dcolumn}% Align table columns on decimal point
\usepackage{bm}% bold math

\newcommand{\Real}{\mathop{\textrm{Re}}}
\begin{document}

\preprint{APS/123-QED}

\title{Closed-form expression for the magnetic shielding constant of the relativistic hydrogenlike atom in an arbitrary discrete energy eigenstate: 
Application of the Sturmian expansion of the generalized 
Dirac--Coulomb Green function}% Force line breaks with \\

\author{Patrycja Stefa{\'n}ska}
 \email{pstefanska@mif.pg.gda.pl}
\affiliation{Atomic Physics Division, \mbox{Department of Atomic, Molecular and Optical Physics,} Faculty of Applied Physics and Mathematics, Gda{\'n}sk University of Technology,  Narutowicza 11/12, 80--233 Gda{\'n}sk, Poland
}%


\begin{abstract}
\begin{center}
\textbf{Published as: Phys.\ Rev.\ A 94 (2016) 012508/1-15}
\\*[1ex]
\textbf{doi: 10.1103/PhysRevA.94.012508}\\*[5ex]
\textbf{Abstract} \\*[0.5ex]
\end{center}
We present analytical derivation of the closed-form expression for the dipole magnetic shielding constant of a Dirac one-electron atom being in an arbitrary discrete energy eigenstate. The external magnetic field, by which the atomic state is perturbed, is assumed to be weak, uniform and time independent. With respect to the atomic nucleus we assume that it is pointlike, spinless, motionless and of charge $Ze$. Calculations are based on the Sturmian expansion of the generalized Dirac--Coulomb Green function [R.\/~Szmytkowski, J.\ Phys.\ B \textbf{30}, 825 (1997); erratum \textbf{30}, 2747 (1997)], combined with the theory of hypergeometric functions. The final result is of an elementary form and agrees with corresponding formulas obtained earlier by other authors for some particular states of the atom.
\end{abstract}

%\keywords{magnetic dipole moment; magnetic field;
%magnetizability; Dirac one-electron atom; Dirac--Coulomb Green
%function; Sturmian functions}%Use showkeys class option if keyword
\maketitle

\section{\label{sec:I}Introduction}
\setcounter{equation}{0}

Almost two decades ago, our group has launched a research program for the systematic calculations of various electromagnetic susceptibilities of Dirac one-electron atoms. As the main goal of the challenge we set ourselves the purely analytical description of the influence of a weak and uniform electric and magnetic field on the atomic state. To provide the calculations of atomic parameters characterizing the response of the relativistic hydrogenlike atom to the external perturbations, one should have a convenient tool that enables us to obtain the closed-form expressions for them. One such useful analytical technique is based on the Sturmian expansion of the generalized Dirac--Coulomb Green function (GDCGF) proposed by Szmytkowski in Ref.\ \cite{Szmy97}. In a series of papers, we have used it to derive the formulas for the static and dynamic electric dipole polarizabilities \cite{Szmy97,Szmy04,Szmy02a}, the induced magnetic anapole moment \cite{Miel06}, the dipole magnetizability \cite{Szmy02b}, the electric and magnetic dipole shielding factors \cite{Stef12,Szmy11} and the cross susceptibilities \cite{Szmy12,Szmy14} of the ground state of hydrogenic atoms. Recently, Szmytkowski and {\L}ukasik extended our previous considerations to the case of multipole perturbing fields \cite{Szmy16,Szmy16b}. Using the same analytical method as in the aforementioned articles, they found the expressions for some families of multipole atomic susceptibilities of the atom in the ground state.  

Some time ago, we proposed another generalization of some of the physical problems described in Refs.\ \cite{Szmy97,Szmy04,Szmy02a,Miel06,Szmy02b,Stef12,Szmy11,Szmy12,Szmy14}. Utilizing the Sturmian series representation of the first-order GDCGF \cite{Szmy97}, we derived analytically the closed-form expressions for the magnetizability \cite{Stef15,Stef16b} and the electric quadrupole moment induced by the dipole magnetic field \cite{Stef16} in an \emph{arbitrary} discrete energy eigenstate of the atom. The final results from Refs.\ \cite{Stef15,Stef16} have the form of the double finite sum of hypergeometric functions ${}_3F_{2}$ with the unit arguments. However, the comparison of their counterparts for the atomic ground state \cite{Szmy02b,Szmy12} (which contain only one ${}_3F_{2}$ function) with the elementary formula describing the dipole magnetic shielding constant provided e.g. in Ref.\ \cite{Stef12}, suggests that such general expression for the later atomic parameter should have a bit simpler form than the corresponding results presented in Refs.\ \cite{Stef15,Stef16}. To check the validity of our suspicions, we set ourselves the goal to find the explicit analytical expression for the magnetic shielding constant ($\sigma$) of the Dirac one-electron atom in an arbitrary excited state. 

The first relativistic studies of that quantity for the atomic ground state have been carried out by Zapryagaev \emph{et al.\/} \cite{Zapr74}  (see also Refs.\ \cite{Zapr81,Zapr85}). The main tool used in their calculations was the Sturmian expansion of the second-order Dirac--Coulomb Green function, proposed a few years earlier \cite{Zon72, Mana73}. The formula for $\sigma$ was rederived in 1999, almost simultaneously, by Moore \cite{Moor99} and Pyper and Zhang \cite{Pype99}, and later by Cheng \emph{et al.\/} \cite{Chen09} and us \cite{Szmy11,Stef12}.  Moore obtained also the expressions for this parameter for states with zero radial quantum number, while Pyper and Zhang found additionally the appropriate formulas for states $2p_{1/2}$ and $2p_{3/2}$. Both analytical methods used in Refs.\ \cite{Moor99,Pype99} were based one the general idea proposed a long time ago by Ramsey \cite{Rams50}, that the total magnetic shielding constant for a given atomic state may be presented as a sum of the paramagnetic and diamagnetic terms. Ten years later, Ivanov \emph{et al.\/} \cite{Ivan09} found the formula for $\sigma$ for the states with the angular-plus-parity symmetry quantum number $\kappa=-1$. In the present article,  we provide a fully relativistic analytical derivation for $\sigma$ of the hydrogenlike atom being in an \emph{arbitrary} discrete energy eigenstate. The final expression for the magnetic shielding constant we arrive at by using the Sturmian expansion of the first-order Dirac--Coulomb Green function \cite{Szmy97}, is of an elementary form and reproduces all previously found formulas for some particular atomic states.

\section{Preliminaries}
\label{II}
\setcounter{equation}{0}
In this work, we shall be concerned with a hydrogenlike atom placed in a weak, static, uniform magnetic field of induction $\boldsymbol{\mathcal{B}}$ directed along the $z$ axis of a Cartesian coordinate system. If the wave function $\Psi(\boldsymbol{r})$ of the electron of mass $m_e$ and charge $-e$ is determined within the framework of Dirac quantum theory, the energy eigenvalue problem for bound states of the system under study is constituted by the Dirac equation
\begin{equation}
\left[
-\mathrm{i}c\hbar\boldsymbol{\alpha}\cdot\boldsymbol{\nabla}
+\frac{1}{2} e c \boldsymbol{\alpha} \cdot \left(\boldsymbol{\mathcal{B}} \times \boldsymbol{r}\right)
+\beta m_ec^{2}-\frac{Ze^{2}}{(4\pi\epsilon_{0})r}-E
\right]
\Psi(\boldsymbol{r})=0
\label{2.1}
\end{equation}
(with $\boldsymbol{\alpha}$ and $\beta$ denoting the standard Dirac matrices \cite{Schi55}), supplemented by the boundary conditions
\begin{equation}
r \Psi(\boldsymbol{r}) \stackrel{r \to 0}{\longrightarrow}0, 
\qquad \qquad 
r^{3/2} \Psi(\boldsymbol{r}) \stackrel{r \to \infty}{\longrightarrow}0. 
\label{2.2}
\end{equation}
In virtue of our assumption that the external magnetic field is weak, the interaction operator $\hat{H}^{(1)}=\frac{1}{2} e c \boldsymbol{\alpha} \cdot \left(\boldsymbol{\mathcal{B}} \times \boldsymbol{r}\right)$, appearing in Eq.\ (\ref{2.1}), may be considered as a small perturbation of the Dirac--Coulomb Hamiltonian for an isolated atom. The corresponding zeroth-order bound-state eigenproblem is given by the equation
\begin{equation}
\left[
-\mathrm{i}c\hbar\boldsymbol{\alpha}\cdot\boldsymbol{\nabla}
+\beta m_e c^{2}-\frac{Ze^{2}}{(4\pi\epsilon_{0})r}-E^{(0)}
\right]
\Psi^{(0)}(\boldsymbol{r})=0,
\label{2.3}
\end{equation}
with the boundary conditions
\begin{equation}
r \Psi^{(0)}(\boldsymbol{r}) \stackrel{r \to 0}{\longrightarrow}0, 
\qquad \qquad 
r^{3/2} \Psi^{(0)}(\boldsymbol{r}) \stackrel{r \to \infty}{\longrightarrow}0. 
\label{2.4}
\end{equation}

If the perturbing field is switched off, the atomic state energy levels are
\begin{equation}
E^{(0)} \equiv E_{n \kappa}^{(0)}=m_ec^2\frac{n+\gamma_{\kappa}}{N_{n \kappa}},
\label{2.5}
\end{equation}
where
\begin{equation}
N_{n\kappa}=\sqrt{n^2+2n\gamma_{\kappa}+\kappa^2}
\label{2.6}
\end{equation}
and
\begin{equation}
\gamma_{\kappa}=\sqrt{\kappa^2-(\alpha Z)^2}.
\label{2.7}
\end{equation}
Here and in the rest of this paper, $n$ denotes the \emph{radial} quantum number, $\kappa$ is the Dirac quantum number which takes integer values different from zero (they are related to the principal quantum number ${N}$ through the formula ${N}=n+|\kappa|$), while $\alpha$ is the Sommerfeld's fine-structure constant (the reader should not confuse it with the Dirac matrix $\boldsymbol{\alpha}$). The eigenfunctions of the Dirac--Coulomb problem (\ref{2.3})--(\ref{2.4}) associated with the eigenvalue $E_{n \kappa}^{(0)}$ are
\begin{equation}
\Psi^{(0)}(\boldsymbol{r}) \equiv \Psi_{n \kappa \mu}^{(0)}(\boldsymbol{r}) 
=\frac{1}{r} 
\left(
\begin{array} {c}
P_{n\kappa}^{(0)}(r) \Omega_{\kappa\mu}(\boldsymbol{n}_r) \\ 
\textrm{i} Q_{n\kappa}^{(0)}(r) \Omega_{-\kappa\mu}(\boldsymbol{n}_r) 
\end{array} 
\right),
\label{2.8}
\end{equation}
where $\Omega_{\kappa \mu}(\boldsymbol{n}_r)$ (with $\boldsymbol{n}_r=\boldsymbol{r}/r$ and $\mu=-|\kappa|+\frac{1}{2}, -|\kappa|+\frac{3}{2}, \ldots , |\kappa|-\frac{1}{2}$) are the orthonormal spherical spinors defined as in Ref.\ \cite{Szmy07} (the space quantization axis is chosen along the direction of an external magnetic field), whereas the radial functions are given by
\begin{equation}
P_{n\kappa}^{(0)}(r)=\mathsf{f}_{n\kappa}\sqrt{1+\epsilon_{n\kappa}}
\left(
\lambda_{n\kappa} r\right)^{\gamma_{\kappa}}\textrm{e}^{-\lambda_{n\kappa} r/2}
\left[
L_{n-1}^{(2\gamma_{\kappa})}\left(\lambda_{n\kappa} r\right)
+\frac{\kappa-N_{n\kappa}}{n+2\gamma_{\kappa}}L_{n}^{(2\gamma_{\kappa})}\left(\lambda_{n\kappa} r\right)
\right],
\label{2.9}
\end{equation}
\begin{equation}
Q_{n\kappa}^{(0)}(r)=\mathsf{f}_{n\kappa}\sqrt{1-\epsilon_{n\kappa}} 
\left(
\lambda_{n\kappa} r\right)^{\gamma_{\kappa}}\textrm{e}^{-\lambda_{n\kappa} r/2}
\left[
L_{n-1}^{(2\gamma_{\kappa})}\left(\lambda_{n\kappa} r\right)
-\frac{\kappa-N_{n\kappa}}{n+2\gamma_{\kappa}}L_{n}^{(2\gamma_{\kappa})}\left(\lambda_{n\kappa} r\right)
\right].
\label{2.10}
\end{equation}
Here and below, $L_{n}^{(\beta)}(\rho)$ is the generalized Laguerre polynomial \cite{Magn66} [it is understood that $L_{-1}^{(2\gamma_{\kappa})}(\rho) \equiv 0$],  
\begin{equation}
\epsilon_{n \kappa}=\frac{E_{n\kappa}^{(0)}}{m_ec^2}=\frac{n+\gamma_{\kappa}}{N_{n\kappa}}, \qquad \lambda_{n\kappa}=\frac{2Z}{a_0 N_{n\kappa}}
\label{2.11}
\end{equation}
(with $a_0$ being the Bohr radius) and 
\begin{equation}
\mathsf{f}_{n\kappa}=\sqrt{\frac{Z}{2a_0} \frac{(n+2\gamma_{\kappa}) n!}{N_{n\kappa}^2(N_{n\kappa}-\kappa)\Gamma(n+2\gamma_{\kappa})}}. 
\label{2.12}
\end{equation}
It is easy to prove the following normalizing relation:
\begin{equation}
\int_0^{\infty} \textrm{d}r 
\left\{
[P_{n\kappa}^{(0)}(r)]^2+[Q_{n\kappa}^{(0)}(r)]^2 
\right\}
=1,
\label{2.13}
\end{equation}
with the use of which one can show that the unperturbed function $\Psi^{(0)}(\boldsymbol{r})$ is normalized to unity in the sense of
\begin{equation}
\int_{\mathbb{R}^3} \textrm{d}^3\boldsymbol{r} \: \Psi^{(0)\dagger}(\boldsymbol{r}) \Psi^{(0)}(\boldsymbol{r})=1
\label{2.14}
\end{equation}
and that the functions from Eq.\ (\ref{2.8}) are adjusted to the perturbation $\hat{H}^{(1)}$; i.e., they diagonalize the matrix of that operator. Thus, to the lowest order in the perturbing field, we can approximate the solutions of eigenproblem (\ref{2.1})--(\ref{2.2}) as
\begin{equation}
\Psi(\boldsymbol{r}) \simeq \Psi^{(0)}(\boldsymbol{r})+\Psi^{(1)}(\boldsymbol{r}) \qquad \textrm{and} \qquad E \simeq E^{(0)}+E^{(1)}.
\label{2.15}
\end{equation}
The corrections $\Psi^{(1)}(\boldsymbol{r})$ and $E^{(1)}$, which are assumed to be small quantities of the first order in the magnetic field strength $\mathcal{B}=|\boldsymbol{\mathcal{B}}|$, solve the inhomogeneous differential equation
\begin{equation}
\left[
-\textrm{i} c \hbar \boldsymbol{\alpha} \cdot \boldsymbol{\nabla} 
+\beta m_ec^2  
-\frac{Z e^2}{(4\pi \epsilon_0)r} 
-E^{(0)} 
\right]
\Psi^{(1)}(\boldsymbol{r})
=- 
\left[
\frac{1}{2} e c \boldsymbol{\mathcal{B}} \cdot 
\left(
\boldsymbol{r} \times \boldsymbol{\alpha}
\right)
-E^{(1)} 
\right] 
\Psi^{(0)}(\boldsymbol{r}), 
\label{2.16}
\end{equation}
with the boundary conditions
\begin{equation}
r \Psi^{(1)}(\boldsymbol{r}) \stackrel{r \to 0}{\longrightarrow}0, 
\qquad \qquad
 r^{3/2} \Psi^{(1)}(\boldsymbol{r}) \stackrel{r \to \infty}{\longrightarrow}0. 
\label{2.17}
\end{equation}
Carrying out standard transformations from the perturbation theory and taking advantage of the constraint
\begin{equation}
\int_{\mathbb{R}^3} \textrm{d}^3\boldsymbol{r}\: \Psi^{(0)\dagger}(\boldsymbol{r}) \Psi^{(1)}(\boldsymbol{r})=0,
\label{2.18}
\end{equation}
we obtain the first-order corrections to the wave function of an electron and to its energy in the form
\begin{equation}
E^{(1)} \equiv E_{n\kappa\mu}^{(1)}=\frac{1}{2} ec \boldsymbol{\mathcal{B}} \cdot \int_{\mathbb{R}^3} \textrm{d}^3\boldsymbol{r} \: \Psi_{n \kappa \mu}^{(0)\dagger}(\boldsymbol{r}) 
\left(
\boldsymbol{r} \times \boldsymbol{\alpha} 
\right) 
\Psi_{n \kappa \mu}^{(0)}(\boldsymbol{r})
\label{2.19}
\end{equation}
and
\begin{eqnarray}
\Psi^{(1)}(\boldsymbol{r}) \equiv \Psi_{n \kappa \mu}^{(1)}(\boldsymbol{r})= 
-\int_{\mathbb{R}^3} \textrm{d}^3\boldsymbol{r}' \: \bar{G}^{(0)}(\boldsymbol{r},\boldsymbol{r}') 
\left[ 
\frac{1}{2} e c \boldsymbol{\mathcal{B}} \cdot 
\left( 
\boldsymbol{r}' \times \boldsymbol{\alpha}
\right)
-E_{n \kappa \mu}^{(1)} 
\right] 
\Psi_{n \kappa \mu}^{(0)}(\boldsymbol{r}'),
\label{2.20}
\end{eqnarray}
in which $\bar{G}^{(0)}(\boldsymbol{r},\boldsymbol{r}')$ is the generalized Dirac--Coulomb Green function associated with the energy level $E_{n\kappa}^{(0)}$ of an isolated atom. The formula in Eq.\ (\ref{2.20}) may be cast into a bit simpler form, if one takes into account the orthogonality condition 
\begin{equation}
\int_{\mathbb{R}^3} \textrm{d}^3\boldsymbol{r} \: \Psi_{n \kappa' \mu}^{(0)\dagger}(\boldsymbol{r}) \bar{G}^{(0)}(\boldsymbol{r},\boldsymbol{r}')=0 
\qquad \quad 
   (\kappa'= \pm \kappa)
\label{2.21}
\end{equation}
and the fact that $\bar{G}^{(0)}(\boldsymbol{r},\boldsymbol{r}')$ is Hermitian in the sense of
\begin{equation}
\bar{G}^{(0)}(\boldsymbol{r},\boldsymbol{r}')=\bar{G}^{(0)\dagger}(\boldsymbol{r}',\boldsymbol{r}).
\label{2.22}
\end{equation}
Then one arrives at the expression
\begin{equation}
\Psi_{n \kappa \mu}^{(1)}(\boldsymbol{r})=-\frac{1}{2} e c \boldsymbol{\mathcal{B}} \cdot \int_{\mathbb{R}^3} \textrm{d}^3\boldsymbol{r'} \:\bar{G}^{(0)}(\boldsymbol{r},\boldsymbol{r}') 
\left( 
\boldsymbol{r}' \times \boldsymbol{\alpha}
\right) 
\Psi_{n \kappa \mu}^{(0)}(\boldsymbol{r}').
\label{2.23}
\end{equation}
For readability, in the rest of this paper we shall omit all the subscripts at $\Psi_{n\kappa \mu}^{(0)}$ and $\Psi_{n\kappa \mu}^{(1)}$.

\section{Analysis of the magnetic field at the location of the atomic nucleus}
\label{III}
\setcounter{equation}{0}

The electric current associated with the electron being in the state $\Psi(\boldsymbol{r})$  can be describe by the density function $\boldsymbol{j}(\boldsymbol{r})$, which --- in the Dirac theory --- is given by
\begin{equation}
\boldsymbol{j}(\boldsymbol{r})=\frac{-e c\Psi^{\dagger}(\boldsymbol{r}) \boldsymbol{\alpha} \Psi(\boldsymbol{r})}{{\int_{\mathbb{R}^3} \textrm{d}^3\boldsymbol{r'} \Psi^{\dagger}(\boldsymbol{r'}) \Psi(\boldsymbol{r'})}}.
\label{3.1}
\end{equation}
Using the Biot--Savart law, we find that at the point $\boldsymbol{r'}$ the magnetic field due to the current distribution $\boldsymbol{j}(\boldsymbol{r})$ is
\begin{equation}
\boldsymbol{B}(\boldsymbol{r'})=\frac{\mu_0}{4\pi} \int_{\mathbb{R}^3} \textrm{d}^3\boldsymbol{r} \: \boldsymbol{j}(\boldsymbol{r}) \times \frac{\boldsymbol{r'}-\boldsymbol{r}}{|\boldsymbol{r'}-\boldsymbol{r}|^3},
\label{3.2}
\end{equation}
where $\mu_0$ is the permeability of vacuum. In the particular case when the observation point is located at the nucleus, i.e., at $\boldsymbol{r'}=\boldsymbol{0}$, the above expression simplifies to
\begin{equation}
\boldsymbol{B}(\boldsymbol{0})=\frac{\mu_0}{4\pi} \int_{\mathbb{R}^3} \textrm{d}^3\boldsymbol{r} \: \frac{\boldsymbol{r} \times \boldsymbol{j}(\boldsymbol{r})}{r^3}.
\label{3.3}
\end{equation}
Since the electronic wave function $\Psi(\boldsymbol{r})$ is known to the first order in the perturbing field $\boldsymbol{\mathcal{B}}$, the current distribution from Eq.\ (\ref{3.1}), with the aid of Eqs.\ (\ref{2.14}), (\ref{2.15}), and (\ref{2.18}),  may be approximated as
\begin{equation}
\boldsymbol{j}(\boldsymbol{r}) \simeq \boldsymbol{j}^{(0)}(\boldsymbol{r})+\boldsymbol{j}^{(1)}(\boldsymbol{r}),
\label{3.4}
\end{equation}
where
\begin{equation}
\boldsymbol{j}^{(0)}(\boldsymbol{r})=-ec\Psi^{(0)\dagger}(\boldsymbol{r}) \boldsymbol{\alpha} \Psi^{(0)}(\boldsymbol{r})  
\label{3.5}
\end{equation}
is the current density in the unperturbed atom, whereas
\begin{equation}
\boldsymbol{j}^{(1)}(\boldsymbol{r})=-2ec\textrm{Re}\left[\Psi^{(0)\dagger}(\boldsymbol{r})\boldsymbol{\alpha} \Psi^{(1)}(\boldsymbol{r})\right]
\label{3.6}
\end{equation}
is the first-order contribution to an induced current density in the perturbed atomic state $\Psi(\boldsymbol{r})$. Consequently, it follows that
 \begin{equation}
 \boldsymbol{B}(\boldsymbol{0}) \simeq \boldsymbol{B}^{(0)}(\boldsymbol{0})+\boldsymbol{B}^{(1)}(\boldsymbol{0}),
 \label{3.7}
\end{equation}
where
\begin{equation}
 \boldsymbol{B}^{(0)}(\boldsymbol{0})=-e c \frac{\mu_0}{4\pi} \int_{\mathbb{R}^3} \textrm{d}^3\boldsymbol{r} \frac{\Psi^{(0)\dagger}(\boldsymbol{r}) (\boldsymbol{r}\times \boldsymbol{\alpha}) \Psi^{(0)}(\boldsymbol{r})}{r^3}
  \label{3.8}
\end{equation}
 is the magnetic field at the location of the nucleus of an isolated atom, while 
 \begin{equation}
 \boldsymbol{B}^{(1)}(\boldsymbol{0})=-2ec \frac{\mu_0}{4\pi} \textrm{Re} \int_{\mathbb{R}^3} \textrm{d}^3\boldsymbol{r} \frac{\Psi^{(0) \dagger}(\boldsymbol{r}) (\boldsymbol{r} \times \boldsymbol{\alpha}) \Psi^{(1)}(\boldsymbol{r})}{r^3}
  \label{3.9}
\end{equation}
describes the change of the magnetic field induced there by the perturbation.

To find $\boldsymbol{B}^{(0)}(\boldsymbol{0})$, we rewrite Eq.\ (\ref{3.8}) in the form of
\begin{equation}
 \boldsymbol{B}^{(0)}(\boldsymbol{0})=-e c \frac{\mu_0}{4\pi} \sum_{j=-1}^1 \boldsymbol{e}_{j}^{*} \int_{\mathbb{R}^3} \textrm{d}^3\boldsymbol{r} \frac{\Psi^{(0) \dagger}(\boldsymbol{r}) \boldsymbol{e}_{j} \cdot (\boldsymbol{n}_{r}\times\boldsymbol{\alpha}) \Psi^{(0)}(\boldsymbol{r})}{r^2},
  \label{3.10}
\end{equation}
where $\boldsymbol{e}_{j}$, with $j=0,\pm1$, are the unit vectors of the cyclic basis, related to the Cartesian unit vectors through
\begin{equation}
\boldsymbol{e}_{0}=\boldsymbol{n}_{z}, \qquad \boldsymbol{e}_{\pm1}=\mp \frac{1}{\sqrt{2}}\left(\boldsymbol{n}_{x} \pm \textrm{i}\boldsymbol{n}_{y}\right).
\label{3.11}
\end{equation}
By substituting $\Psi^{(0)}(\boldsymbol{r})$, as given by Eq.\ (\ref{2.8}), using the following spherical spinor identities \cite{Szmy07} 
\begin{equation}
\boldsymbol{e}_0 \cdot \left(\boldsymbol{n}_r \times \boldsymbol{\sigma} \right) \Omega_{\kappa \mu}(\boldsymbol{n}_r)=\textrm{i}\frac{4\mu \kappa}{4\kappa^2-1} \Omega_{-\kappa \mu}(\boldsymbol{n}_r)+\textrm{i}\frac{\sqrt{\left(\kappa+\frac{1}{2} \right)^2-\mu^2}}{|2\kappa+1|} \Omega_{\kappa+1, \mu}(\boldsymbol{n}_r)
-\textrm{i} \frac{\sqrt{\left(\kappa-\frac{1}{2}\right)^2-\mu^2}}{|2\kappa-1|} \Omega_{\kappa-1, \mu}(\boldsymbol{n}_r),
\label{3.12}
\end{equation} 
\begin{eqnarray}
\boldsymbol{e}_{\pm 1}\cdot \left(\boldsymbol{n}_r \times \boldsymbol{\sigma} \right) \Omega_{\kappa \mu}(\boldsymbol{n}_r)&=&\mp \textrm{i} 2\sqrt{2}\kappa\frac{\sqrt{\kappa^2-\left(\mu \pm \frac{1}{2}\right)^2}}{4\kappa^2-1} \Omega_{-\kappa, \mu \pm 1}(\boldsymbol{n}_r)
+\textrm{i} \frac{\sqrt{\left(\kappa \pm \mu +\frac{1}{2}\right)\left(\kappa \pm \mu +\frac{3}{2}\right)}}{\sqrt{2} (2\kappa+1)} \Omega_{\kappa+1, \mu \pm 1}(\boldsymbol{n}_r){}
\nonumber
\\
&&+
\textrm{i} \frac{\sqrt{\left(\kappa \mp \mu -\frac{1}{2}\right)\left(\kappa \mp \mu -\frac{3}{2}\right)}}{\sqrt{2}(2\kappa-1)} \Omega_{\kappa-1, \mu \pm 1}(\boldsymbol{n}_r),
\label{3.13}
\end{eqnarray}
then carrying out integration over the angular variables, with the help of the orthonormal relation
\begin{equation}
\oint_{4\pi} \textrm{d}^2\boldsymbol{n}_r \: \Omega_{\kappa\mu}^{\dagger}(\boldsymbol{n}_r) \: \Omega_{\kappa'\mu'}(\boldsymbol{n}_r)=\delta_{\kappa \kappa'}\delta_{\mu \mu'},
\label{3.14}
\end{equation}
we arrive at
\begin{equation}
 \boldsymbol{B}^{(0)}(\boldsymbol{0})=-e c \frac{\mu_0}{4\pi} \frac{8\kappa\mu}{4\kappa^2-1} \, \boldsymbol{n}_{z} \int_0^{\infty} \textrm{d}r \: r^{-2} P_{n\kappa}^{(0)}(r)Q_{n\kappa}^{(0)}(r) .
\label{3.15}
\end{equation}
By utilizing Eqs.\ (\ref{2.9}) and (\ref{2.10}) and the relation
\begin{equation}
\int_0^{\infty} \textrm{d}\rho \: \rho^{\beta-2} \textrm{e}^{-\rho} L_{m}^{(\beta)}(\rho) \: L_{n}^{(\beta)}(\rho) 
= \frac{\Gamma(\beta+n+1)}{n! \beta\left(\beta^2-1\right)}\left[\beta(m-n+1)+(m+n+1)\right] \quad \quad \left(\beta>1, \ n \leqslant m\right),
\label{3.16}
\end{equation}
taking advantage of Eqs.\ (\ref{2.11}) and (\ref{2.12}), one finds that 
\begin{equation}
\int_0^{\infty} \textrm{d}r \: r^{-2} P_{n\kappa}^{(0)}(r)Q_{n\kappa}^{(0)}(r)=\frac{\alpha Z^3}{a_0^2}\frac{2\kappa(n+\gamma_{\kappa})-N_{n\kappa}}{\gamma_{\kappa}(4\gamma_{\kappa}^2-1)N_{n\kappa}^4}.
\label{3.17}
\end{equation}
(Notice, that for the states with $\kappa=\pm1$, the above integral converges at its lower limit provided $\gamma_1>\frac{1}{2}$, which constrains the nuclear charge as $Z<\alpha^{-1}\frac{\sqrt{3}}{2} \simeq 118.67$.) Consequently, for an isolated atom we have 
 \begin{equation}
 \boldsymbol{B}^{(0)}(\boldsymbol{0})= 
\frac{16 \kappa \mu}{4\kappa^2-1}
\frac{N_{n \kappa}-2\kappa(n+\gamma_{\kappa})}{\gamma_{\kappa}(4\gamma_{\kappa}^2-1)}
\frac{Z^3}{N_{n\kappa}^4}b_0 \: \boldsymbol{n}_z,
\label{3.18}
 \end{equation} 
where 
\begin{equation}
b_0=\frac{\mu_0}{4\pi} \frac{\mu_B}{a_0^3}=\frac{\alpha^2\hbar}{2ea_0^2}\simeq 6.26 \: \textrm{T}
\label{3.19}
\end{equation}
is the atomic unit of the magnetic induction ($\mu_B$ is the Bohr magneton).

To evaluate the first-order approximation to the magnetic field induced at the nucleus, we plug Eq.\ (\ref{2.23}) into Eq.\ (\ref{3.9}). This gives
 \begin{equation}
\boldsymbol{B}^{(1)}(\boldsymbol{0})=\frac{e^2}{4\pi\epsilon_0} \textrm{Re}\int_{\mathbb{R}^3}\textrm{d}^3 \boldsymbol{r}\int_{\mathbb{R}^3}\textrm{d}^3 \boldsymbol{r'} \: r^{-3}\Psi^{(0)\dagger}(\boldsymbol{r})\left(\boldsymbol{r}\times\boldsymbol{\alpha}\right) \bar{G}^{(0)}(\boldsymbol{r}, \boldsymbol{r'}) \boldsymbol{\mathcal{B}} \cdot (\boldsymbol{r'}\times\boldsymbol{\alpha})\Psi^{(0)}(\boldsymbol{r'}),
\label{3.20}
\end{equation}
where we have used the Maxwell's identity $\mu_0\epsilon_0c^2=1$. Now, keeping in mind the direction of the perturbing field, i.e., $\boldsymbol{\mathcal{B}}=\mathcal{B}\boldsymbol{n}_{z}=\mathcal{B}\boldsymbol{e}_{0}$, the above expression may be conveniently rewritten as
 \begin{equation}
\boldsymbol{B}^{(1)}(\boldsymbol{0})=\frac{e^2 \mathcal{B}}{4\pi\epsilon_0}
\textrm{Re}\sum_{j=-1}^{+1} \boldsymbol{e}_{j}
\int_{\mathbb{R}^3}\textrm{d}^3 \boldsymbol{r}\int_{\mathbb{R}^3}\textrm{d}^3 \boldsymbol{r'} 
\left[
\boldsymbol{e}_{j}\cdot
\left(
\boldsymbol{n}_{r}\times\boldsymbol{\alpha}
\right)
\Psi^{(0)}(\boldsymbol{r})\right]^{\dagger} 
r^{-2}\bar{G}^{(0)}(\boldsymbol{r},\boldsymbol{r'}) r'
\left[
\boldsymbol{e}_{0}\cdot
\left(
\boldsymbol{n}_{r}'\times\boldsymbol{\alpha}
\right)
\Psi^{(0)}(\boldsymbol{r'})\right].
\label{3.21}
\end{equation}
In the next step, we plug Eq.\ (\ref{2.8}) into Eq.\ (\ref{3.21}), and invoke the partial-wave expansion of the generalized Dirac--Coulomb Green function
{\footnotesize
\begin{eqnarray}
\bar{G}\mbox{}^{(0)}(\boldsymbol{r},\boldsymbol{r}')
= \frac{4\pi\epsilon_{0}}{e^{2}} 
\sum_{\substack{\kappa'=-\infty \\ (\kappa'\neq0)}}^{\infty}
\sum_{\mu'=-|\kappa'|+1/2}^{|\kappa'|-1/2}\frac{1}{rr'} 
\left(
\begin{array}{cc}
\bar{g}\mbox{}^{(0)}_{(++)\kappa'}(r,r')
\Omega_{\kappa' \mu'}(\boldsymbol{n}_{r})
\Omega_{\kappa' \mu'}^{\dag}(\boldsymbol{n}_{r}^{\prime}) &
-\mathrm{i}\bar{g}\mbox{}^{(0)}_{(+-)\kappa'}(r,r')
\Omega_{\kappa' \mu'}(\boldsymbol{n}_{r})
\Omega_{-\kappa' \mu'}^{\dag}(\boldsymbol{n}_{r}^{\prime}) \\
\mathrm{i}\bar{g}\mbox{}^{(0)}_{(-+)\kappa'}(r,r')
\Omega_{-\kappa' \mu'}(\boldsymbol{n}_{r})
\Omega_{\kappa' \mu'}^{\dag}(\boldsymbol{n}_{r}^{\prime}) &
\bar{g}\mbox{}^{(0)}_{(--)\kappa'}(r,r')
\Omega_{-\kappa' \mu'}(\boldsymbol{n}_{r})
\Omega_{-\kappa' \mu'}^{\dag}(\boldsymbol{n}_{r}^{\prime}) 
\end{array} 
\right).
\label{3.22}
\end{eqnarray}}%
Then, using Eqs.\ (\ref{3.11})--(\ref{3.14}) to carry out integrations over the angles of the vectors $\boldsymbol{r}$ and $\boldsymbol{r'}$, after some labor, we discover that the only nonzero constituent from the sum in Eq.\ (\ref{3.21}) is the one with $j=0$. Hence,  
\begin{equation}
\boldsymbol{B}^{(1)}(\boldsymbol{0})=-\sigma\boldsymbol{\mathcal{B}},
\label{3.23}
\end{equation}
where the proportionality factor $\sigma$ is the magnetic dipole shielding constant for the system under study, given by
\begin{eqnarray}
\sigma&=&\sum_{\substack{\kappa'=-\infty \\ (\kappa'\neq0)}}^{\infty} 
\int_0^{\infty} \textrm{d}r \int_0^{\infty} \textrm{d}r'  
\left( 
\begin{array}{cc} 
Q_{n\kappa}^{(0)}(r) & P_{n\kappa}^{(0)}(r)
\end{array} 
\right) 
 r^{-2}\bar{\mathsf{G}}_{\kappa'}^{(0)}(r,r') r' 
\left( 
\begin{array}{c}
Q_{n\kappa}^{(0)}(r')\\
P_{n\kappa}^{(0)}(r')
\end{array} 
\right) 
{}\nonumber\\
&& \times  
\left[ 
\frac{-16\kappa^2\mu^2}{\left(4\kappa^2-1\right)^2}\delta_{\kappa',\kappa} 
+\frac{\mu^2-\left(\kappa+\frac{1}{2}\right)^2}{\left(2\kappa+1\right)^2}\delta_{\kappa',-\kappa-1}
+\frac{\mu^2-\left(\kappa-\frac{1}{2}\right)^2}{\left(2\kappa-1\right)^2} \delta_{\kappa',-\kappa+1} 
\right].
\label{3.24}
\end{eqnarray}
In the above definition 
\begin{equation}
\bar{\mathsf{G}}\mbox{}^{(0)}_{\kappa'}(r,r')
=\left(
\begin{array}{cc}
\bar{g}\mbox{}^{(0)}_{(++)\kappa'}(r,r') &
\bar{g}\mbox{}^{(0)}_{(+-)\kappa'}(r,r') \\*[1ex]
\bar{g}\mbox{}^{(0)}_{(-+)\kappa'}(r,r') &
\bar{g}\mbox{}^{(0)}_{(--)\kappa'}(r,r')
\end{array}
\right)
\label{3.25}
\end{equation}
is the radial generalized Dirac--Coulomb Green function associated with the angular-plus-parity symmetry quantum number $\kappa'$. Evaluation of the closed-form expression for $\sigma$ will be discussed in details in the next section.

\section{Magnetic shielding constant}
\label{IV}
\setcounter{equation}{0}

Let us rewrite Eq.\ (\ref{3.24}) in the following form:
\begin{equation}
\sigma=\sigma_{\kappa}+\sigma_{-\kappa-1}+\sigma_{-\kappa+1},
\label{4.1}
\end{equation}
where each component corresponds to the particular Kronecker's delta from the former equation. To determine all the three constituents of $\sigma$, due to the expression in Eq.\ (\ref{3.24}), we need to have an explicit formula for the radial generalized Dirac--Coulomb Green function. We shall use its expansions in the Sturmian basis, which are \cite{Szmy97}
\begin{equation}
\bar{\mathsf{G}}_{\kappa'}^{(0)}(r,r')
=
\sum_{n'=-\infty}^{\infty}{\frac{1}{\mu_{n'\kappa'}^{(0)}-1}}
\left( 
\begin{array}{c}
S_{n'\kappa'}^{(0)}(r) \\
T_{n'\kappa'}^{(0)}(r)
\end{array} 
\right) 
\left(
\begin{array}{cc} 
\mu_{n'\kappa'}^{(0)}S_{n'\kappa'}^{(0)}(r') & 
T_{n'\kappa'}^{(0)}(r')
\end{array}
\right) \qquad   \left(\textrm{for} \quad \kappa' \neq \kappa\right)
\label{4.2}
\end{equation}
and
\begin{eqnarray}
\bar{\mathsf{G}}_{\kappa}^{(0)}(r,r')
&=&
\sum_{\substack{n'=-\infty\\(n'\neq n)}}^{\infty}{\frac{1}{\mu_{n' \kappa}^{(0)}-1}}
\left( 
\begin{array}{c}
S_{n'\kappa}^{(0)}(r) \\
T_{n'\kappa}^{(0)}(r)
\end{array} 
\right)
\left( 
\begin{array}{cc} 
\mu_{n'\kappa}^{(0)} S_{n'\kappa}^{(0)}(r') & 
T_{n'\kappa}^{(0)}(r')\end{array}\right)
 +
\left(
\epsilon_{n\kappa}-\frac{1}{2} 
\right)
\left( 
\begin{array}{c}
S_{n\kappa}^{(0)}(r) \\
T_{n\kappa}^{(0)}(r)
\end{array} 
\right)
\left( 
\begin{array}{cc} 
S_{n\kappa}^{(0)}(r') & 
T_{n\kappa}^{(0)}(r') 
\end{array} 
\right) 
 \nonumber \\
&&+
\left( 
\begin{array}{c}
I_{n\kappa}^{(0)}(r) \\
K_{n\kappa}^{(0)}(r)
\end{array} 
\right)
\left( 
\begin{array}{cc} 
S_{n\kappa}^{(0)}(r') &  
T_{n\kappa}^{(0)}(r') 
\end{array} \right)
 +
\left( 
\begin{array}{c}
S_{n\kappa}^{(0)}(r) \\
T_{n\kappa}^{(0)}(r)
\end{array} 
\right)
\left( 
\begin{array}{cc} 
J_{n\kappa}^{(0)}(r') & 
K_{n\kappa}^{(0)}(r')
\end{array} 
\right) \qquad   \left(\textrm{for} \quad \kappa'=\kappa\right).
\label{4.3}
\end{eqnarray}
In the last two formulas 
\begin{equation}
S_{n'\kappa'}^{(0)}(r)=\mathsf{g}_{n'\kappa'}\sqrt{1+\epsilon_{n\kappa}}
\left(\lambda_{n\kappa} r\right)^{\gamma_{\kappa'}} 
\textrm{e}^{-\lambda_{n\kappa} r/2}  
\left[
L_{|n'|-1}^{(2\gamma_{\kappa'})}\left(\lambda_{n\kappa} r\right) 
+\frac{\kappa'-N_{n'\kappa'}}{|n'|+2\gamma_{\kappa'}}L_{|n'|}^{(2\gamma_{\kappa'})}\left(\lambda_{n\kappa} r\right)
\right]
\label{4.4}
\end{equation}
and
\begin{equation}
T_{n'\kappa'}^{(0)}(r)=\mathsf{g}_{n'\kappa'}\sqrt{1-\epsilon_{n\kappa}}
\left(\lambda_{n\kappa} r\right)^{\gamma_{\kappa'}} 
\textrm{e}^{-\lambda_{n\kappa} r/2}  
\left[
L_{|n'|-1}^{(2\gamma_{\kappa'})}\left(\lambda_{n\kappa} r\right) 
-\frac{\kappa'-N_{n'\kappa'}}{|n'|+2\gamma_{\kappa'}}L_{|n'|}^{(2\gamma_{\kappa'})}\left(\lambda_{n\kappa} r\right)
\right],
\label{4.5}
\end{equation}
with
\begin{equation}
\mathsf{g}_{n'\kappa'}=\sqrt{\frac{N_{n \kappa}(|n'|+2\gamma_{\kappa'})|n'|!}{2 Z N_{n'\kappa'}(N_{n'\kappa'}-\kappa')\Gamma(|n'|+2\gamma_{\kappa'})}},  
\label{4.6}
\end{equation}
are the radial Dirac--Coulomb Sturmian functions associated with the hydrogenic discrete state energy level $E_{n\kappa}^{(0)}$. The functions $I_{n\kappa}^{(0)}(r)$, $J_{n\kappa}^{(0)}(r)$, and $K_{n\kappa}^{(0)}(r)$, also appearing in Eqs.\ (\ref{4.2})--(\ref{4.3}), are defined as \cite{Szmy97} 
\begin{equation}
I_{n\kappa}^{(0)}(r)
=\epsilon_{n \kappa} 
\left[
-\omega_{n\kappa}^{(+)}S_{n\kappa}^{(0)}(r)+\xi_{n\kappa}^{(+)}(r)T_{n\kappa}^{(0)}(r)
\right], 
\label{4.7}
\end{equation}
\begin{equation}
J_{n\kappa}^{(0)}(r)
=\epsilon_{n \kappa} 
\left[
-\omega_{n\kappa}^{(-)}S_{n\kappa}^{(0)}(r)+\xi_{n\kappa}^{(+)}(r)T_{n\kappa}^{(0)}(r)
\right], 
\label{4.8}
\end{equation}
\begin{equation}
K_{n\kappa}^{(0)}(r)
=\epsilon_{n \kappa} 
\left[
\xi_{n\kappa}^{(-)}(r)S_{n\kappa}^{(0)}(r)+\omega_{n\kappa}^{(+)}T_{n\kappa}^{(0)}(r)
\right], 
\label{4.9}
\end{equation}
where $\omega_{n\kappa}^{(\pm)}=\kappa \pm (2\epsilon_{n\kappa})^{-1}$ and $\xi_{n\kappa}^{(\pm)}(r)=m_{e} c (1 \pm \epsilon_{n\kappa}) r/\hbar \pm \alpha Z$. Moreover, the Sturmian eigenvalue occurring in Eqs.\ (\ref{4.2})--(\ref{4.3}) is given by
\begin{equation}
\mu_{n'\kappa'}^{(0)}=\frac{|n'|+\gamma_{\kappa'}+N_{n'\kappa'}}{n+\gamma_{\kappa}+N_{n \kappa}},
\label{4.10}
\end{equation}
with
\begin{equation}
N_{n'\kappa'}
=\pm \sqrt{|n'|+2|n'|\gamma_{\kappa'}+\kappa'^2}
\label{4.11}
\end{equation}
denoting a so-called apparent principal quantum number, which assumes the positive values if $n'>0$ and negative if $n'<0$; when $n'=0$, in the definition (\ref{4.11}) one should choose the plus sign for $\kappa'<0$ and the minus sign for $\kappa'>0$.

After these preparatory steps, we shall concern with the first component of $\sigma$ appearing on the right-hand side of Eq.\ (\ref{4.1}). In view of Eqs.\ (\ref{3.24}) and (\ref{4.3}), $\sigma_{\kappa}$ may be written in the form
\begin{equation}
\sigma_{\kappa}=-\frac{16\kappa^2\mu^2}{(4\kappa^2-1)^2}\left[
\mathcal{I}_{\kappa}^{(\infty)}+\mathcal{I}_{\kappa}^{(a)}+\mathcal{I}_{\kappa}^{(b)}+\mathcal{I}_{\kappa}^{(c)} 
\right],
\label{4.12}
\end{equation}
where 
\begin{eqnarray}
\mathcal{I}_{\kappa}^{(\infty)}=\sum_{\substack{{n'=-\infty}\\(n' \neq n)}}^{\infty}{\frac{1}{\mu_{n' \kappa}^{(0)}-1}} 
\int_0^{\infty} \textrm{d}r \: r^{-2}  
\left[
Q_{n\kappa}^{(0)}(r) S_{n'\kappa}^{(0)}(r)+ P_{n\kappa}^{(0)}(r) T_{n'\kappa}^{(0)}(r)
\right] 
\nonumber \\
\times
\int_0^{\infty} \textrm{d}r' \: r' 
\left[
\mu_{n' \kappa}^{(0)} Q_{n\kappa}^{(0)}(r') S_{n'\kappa}^{(0)}(r')+ P_{n\kappa}^{(0)}(r') T_{n'\kappa}^{(0)}(r')
\right],
\label{4.13}
\end{eqnarray}
\begin{eqnarray}
\mathcal{I}_{\kappa}^{(a)}= 
\left(\epsilon_{n\kappa}-\frac{1}{2} \right) 
\int_0^{\infty} \textrm{d}r \: r^{-2} 
\left[
Q_{n\kappa}^{(0)}(r) S_{n\kappa}^{(0)}(r)+ P_{n\kappa}^{(0)}(r) T_{n\kappa}^{(0)}(r)
\right] 
\int_0^{\infty} \textrm{d}r' \: r' 
\left[
Q_{n\kappa}^{(0)}(r') S_{n\kappa}^{(0)}(r')+ P_{n\kappa}^{(0)}(r') T_{n\kappa}^{(0)}(r')
\right],
\nonumber \\ 
\label{4.14}
\end{eqnarray}
\begin{eqnarray}
\mathcal{I}_{\kappa}^{(b)}=  
\int_0^{\infty} \textrm{d}r \: r^{-2} 
\left[
Q_{n\kappa}^{(0)}(r) I_{n\kappa}^{(0)}(r)+ P_{n\kappa}^{(0)}(r) K_{n\kappa}^{(0)}(r)
\right]
\int_0^{\infty} \textrm{d}r' \: r' 
\left[
Q_{n\kappa}^{(0)}(r') S_{n\kappa}^{(0)}(r')+ P_{n\kappa}^{(0)}(r') T_{n\kappa}^{(0)}(r')
\right], 
\label{4.15}
\end{eqnarray}
\begin{eqnarray}
\mathcal{I}_{\kappa}^{(c)}=
\int_0^{\infty} \textrm{d}r \: r^{-2} 
\left[
Q_{n\kappa}^{(0)}(r) S_{n\kappa}^{(0)}(r)+ P_{n\kappa}^{(0)}(r) T_{n\kappa}^{(0)}(r)
\right] 
\int_0^{\infty} \textrm{d}r' \: r' 
\left[
Q_{n\kappa}^{(0)}(r') J_{n\kappa}^{(0)}(r')+ P_{n\kappa}^{(0)}(r') K_{n\kappa}^{(0)}(r')
\right]. 
\label{4.16}
\end{eqnarray}
Making use of Eqs.\ (\ref{4.7})--(\ref{4.9}) and the relations
\begin{equation}
S_{n\kappa}^{(0)}(r)=\frac{\sqrt{a_0}N_{n\kappa}}{Z}P_{n\kappa}^{(0)}(r), \qquad T_{n\kappa}^{(0)}(r)=\frac{\sqrt{a_0}N_{n\kappa}}{Z}Q_{n\kappa}^{(0)}(r),
\label{4.17}
\end{equation}
after some simplifications, one has
\begin{equation}
\mathcal{I}_{\kappa}^{(a)}+\mathcal{I}_{\kappa}^{(c)}=0
\label{4.18}
\end{equation}
and 
\begin{equation}
\mathcal{I}_{\kappa}^{(b)}=2\epsilon_{n \kappa}\frac{a_0 N_{n\kappa}^2}{Z^2}
\int_0^{\infty} \textrm{d}r \: r^{-2} P_{n\kappa}^{(0)}(r)Q_{n\kappa}^{(0)}(r) \int_0^{\infty} \textrm{d}r' \: r' P_{n\kappa}^{(0)}(r')Q_{n\kappa}^{(0)}(r').
\label{4.19}
\end{equation}
The first radial integral from the above equation has been evaluated earlier; it is given by Eq.\ (\ref{3.17}). To tackle the second one, we utilize Eqs.\ (\ref{2.9})--(\ref{2.10}) and employ the recurrence formula for the Laguerre polynomials \cite[Eq.\ (8.971.5)]{Grad94}
\begin{equation}
L_n^{(\beta)}(\rho)=L_n^{(\beta+1)}(\rho)-L_{n-1}^{(\beta+1)}(\rho)
\label{4.20}
\end{equation}
and the orthogonality relation \cite[Eq.\ (7.414.3)]{Grad94} 
\begin{equation}
\int_0^{\infty} \textrm{d}\rho \: \rho^{\beta}\textrm{e}^{-\rho}L_m^{(\beta)}(\rho)L_n^{(\beta)}(\rho) 
=\frac{\Gamma(n+\beta+1)}{n!}\delta_{m n} 
\quad \quad 
[\Real \beta>-1].
\label{4.21}
\end{equation}
Using also Eqs.\ (\ref{2.11}) and (\ref{2.12}), with some labor, we obtain
\begin{equation}
\int_0^{\infty} \textrm{d}r' \: r' P_{n\kappa}^{(0)}(r')Q_{n\kappa}^{(0)}(r')
=\frac{\alpha a_0}{4N_{n\kappa}}
\left[
2\kappa(n+\gamma_{\kappa})-N_{n\kappa}
\right].
\label{4.22}
\end{equation}
By substituting Eqs.\ (\ref{2.11}), (\ref{3.17})  and (\ref{4.22}) into Eq.\ (\ref{4.19}), we get
\begin{equation}
\mathcal{I}_{\kappa}^{(b)}=\frac{\alpha^2 Z}{N_{n\kappa}^4} 
\frac{(n+\gamma_{\kappa})
\left[
2\kappa(n+\gamma_{\kappa})-N_{n \kappa} 
\right]^2}
{2\gamma_{\kappa}(4\gamma_{\kappa}^2-1)}.
\label{4.23}
\end{equation}
To complete our derivation of $\sigma_{\kappa}$, we have to find the expression for the remaining component from Eq.\ (\ref{4.12}), namely  $\mathcal{I}_{\kappa}^{(\infty)}$. To accomplish the goal, we plug Eqs.\ (\ref{2.9})--(\ref{2.12}), (\ref{4.4})--(\ref{4.6}), and (\ref{4.10})--(\ref{4.11}) into Eq.\ (\ref{4.13}), and exploit the relations (\ref{3.16}), (\ref{4.20}), and (\ref{4.21}). After some rearrangements, we find
\begin{equation}
\mathcal{I}_{\kappa}^{(\infty)}
=\frac{\alpha^2 Z}{8\gamma_{\kappa}(4\gamma_{\kappa}^2-1)N_{n\kappa}^2} 
\sum_{\substack{{n'=-\infty}\\(n' \neq n)}}^{\infty}\frac{1}{N_{n' \kappa}} 
\sum_{i=-2}^{+2} \mathsf{A}_{n'}^{(i)}\delta_{|n'|,n+i}, 
\label{4.24}
\end{equation}
with the coefficients
\begin{equation}
\mathsf{A}_{n'}^{(0)}=8n(n+2\gamma_{\kappa})\epsilon_{n\kappa}\left[N_{n'\kappa}^2+2(n+\gamma_{\kappa})^2\right],
\label{4.25}
\end{equation}
\begin{equation}
\mathsf{A}_{n'}^{(\pm 1)}=4(N_{n\kappa} \mp \kappa)(N_{n'\kappa} \pm \kappa)\left(2\kappa \pm N_{n\kappa} \mp N_{n'\kappa}\right) \left[2n+2\gamma_{\kappa} \pm 1 \mp \kappa\left(N_{n\kappa}+N_{n'\kappa}\right) \right],
\label{4.26}
\end{equation}
\begin{equation}
\mathsf{A}_{n'}^{(\pm 2)}=(n \pm 1)\left[(n \pm 2)(2n+6\gamma_{\kappa} \pm 3)-n(2n+6\gamma_{\kappa} \pm 1)(N_{n'\kappa} \pm \kappa)\right].
\label{4.27}
\end{equation}
By putting Eqs.\ (\ref{4.18}) and (\ref{4.23})--(\ref{4.27}) into Eq.\ (\ref{4.12}), providing some further simplifications, we arrive at the following expression for the first component of the magnetic shielding constant:
\begin{equation}
\sigma_{\kappa}= \frac{\alpha^2 Z}{N_{n \kappa}^2}  \frac{8\kappa^2 \mu^2}{(4\kappa^2-1)^2}+\frac{\alpha^4 Z^3}{N_{n \kappa}^4}  \frac{32 \kappa^3 \mu^2 \left[2\kappa(n+\gamma_{\kappa})-N_{n \kappa} \right]}{(4\kappa^2-1)^2\gamma_{\kappa}(4\gamma_{\kappa}^2-1)}.
\label{4.28}
\end{equation}
For the atomic ground state, i.e. when $n=0$, $\kappa=-1$, and $\mu=\pm 1/2$, the above result takes the form
\begin{equation}
\sigma_{-1,g}=-\frac{2 \alpha^2 Z}{9} \frac{2\gamma_1^2+\gamma_1-4}{\gamma_1(2\gamma_1-1)}
\label{4.29}
\end{equation}
(the subscript $g$ denotes the ground state), in agreement with the corresponding formula derived by Cheng \emph{et al.} in Ref. \cite{Chen09}.

Now, we shall focus our interests on the two remaining terms of the susceptibility under study, i.e., $\sigma_{-\kappa-1}$ and $\sigma_{-\kappa+1}$. Since the Sturmian expansion of the generalized radial Green function, which we should use in both of these cases, has the same form, given by Eq.\ (\ref{4.2}), it will be convenient to derive expressions for them collectively. According to the formulas in Eqs.\ (\ref{3.24}), (\ref{3.25}), (\ref{4.1}), and (\ref{4.2}), their sum may be written as
\begin{eqnarray}
\sigma_{-\kappa-1}+\sigma_{-\kappa+1}=\sum_{\substack{\kappa'=-\infty \\ (\kappa'\neq0)}}^{\infty}  
\left\{
\left[\frac{\mu^2}{(2\kappa+1)^2}-\frac{1}{4}\right] \delta_{\kappa',-\kappa-1}  
+\left[\frac{\mu^2}{(2\kappa-1)^2}-\frac{1}{4}\right] \delta_{\kappa',-\kappa+1}  
\right\} 
\mathsf{R}_{\kappa'}, 
\label{4.30}
\end{eqnarray}
where
\begin{eqnarray}
\mathsf{R}_{\kappa'}=\sum_{n'=-\infty}^{\infty}\frac{1}{\mu_{n' \kappa'}^{(0)}-1}   
\int_0^{\infty}\textrm{d}r \: r^{-2}
\left[
Q_{n\kappa}^{(0)}(r)S_{n'\kappa'}^{(0)}(r)+P_{n\kappa}^{(0)}(r)T_{n'\kappa'}^{(0)}(r) 
\right] 
\nonumber \\
\times 
\int_0^{\infty} \textrm{d}r' \: r'
\left[
\mu_{n' \kappa'}^{(0)}Q_{n\kappa}^{(0)}(r') S_{n'\kappa'}^{(0)}(r')+P_{n\kappa}^{(0)}(r')T_{n'\kappa'}^{(0)}(r') 
\right].
\label{4.31}
\end{eqnarray}
By exploiting Eqs.\ (\ref{4.4})--(\ref{4.5}) and (\ref{2.9})--(\ref{2.10}) to the first radial integral appearing above and rewriting the Laguerre polynomials from these latter in the form of
\begin{equation}
L_n^{(\beta)}(\rho)=\sum_{k=0}^{n} \frac{(-)^k}{k!} 
\left( 
\begin{array}{c} 
n+\beta \\
n-k 
\end{array} 
\right) 
\rho^k,
\label{4.32}
\end{equation}
after transformation of the integration variable according to $x=\lambda_{n\kappa}r$, we obtain
\begin{eqnarray}
\int_0^{\infty} \textrm{d}r \: r^{-2}
\left[
Q_{n\kappa}^{(0)}(r) S_{n'\kappa'}^{(0)}(r)+P_{n\kappa}^{(0)}(r) T_{n'\kappa'}^{(0)}(r)
\right]
=2\alpha Z\frac{\lambda_{n\kappa}}{N_{n\kappa}}\mathsf{f}_{n\kappa}\mathsf{g}_{n\kappa}\Gamma(n+2\gamma_{\kappa})\sum_{k=0}^{n}\frac{(-)^k}{k!(n-k)!\Gamma(k+2\gamma_{\kappa}+1)}
\nonumber \\
\times
\int_0^{\infty}\textrm{d}x \: x^{\gamma_{\kappa}+\gamma_{\kappa'}+k-2} \textrm{e}^{-x} 
\left[
(n-k)L_{|n'|-1}^{(2\gamma_{\kappa'})}(x)
-\frac{(N_{n\kappa}-\kappa)(N_{n'\kappa'}-\kappa')}{|n'|+2\gamma_{\kappa'}}L_{|n'|}^{(2\gamma_{\kappa'})}(x)
\right].
\label{4.33}
\end{eqnarray}
By employing the relation \cite[Eq.\ (7.414.11)]{Grad94}
 \begin{equation}
\int_0^{\infty} \textrm{d}\rho \: \rho^{\gamma}\textrm{e}^{-\rho}L_n^{(\beta)}(\rho)
=\frac{\Gamma(\gamma+1)\Gamma(n+\beta-\gamma)}{n! \Gamma(\beta-\gamma)}
\qquad 
[\Real(\gamma)>-1],
\label{4.34}
\end{equation}
making use of Eq.\ (\ref{4.11}) to carry out further simplifications, we find that
\begin{eqnarray}
\int_0^{\infty} \textrm{d}r \: r^{-2}
\left[
Q_{n\kappa}^{(0)}(r) S_{n'\kappa'}^{(0)}(r)+P_{n\kappa}^{(0)}(r) T_{n'\kappa'}^{(0)}(r)
\right]
=\frac{4\alpha^2 Z}{a_0}\frac{\mathsf{f}_{n\kappa}\mathsf{g}_{n\kappa}\Gamma(n+2\gamma_{\kappa})}{(|n'|-1)!(N_{n'\kappa'}+\kappa')N_{n\kappa}^2} 
\sum_{k=0}^{n}\frac{(-)^k}{k!}\frac{\Gamma(\gamma_{\kappa}+\gamma_{\kappa'}+k-1)}{\Gamma(\gamma_{\kappa'}-\gamma_{\kappa}-k+2)} 
\nonumber \\
\times  
\frac{\Gamma(|n'|+\gamma_{\kappa'}-\gamma_{\kappa}-k+1)}{(n-k)!\Gamma(k+2\gamma_{\kappa}+1)} 
\left[
(n-k)(N_{n'\kappa'}+\kappa')+(\kappa-N_{n \kappa})(|n'|+\gamma_{\kappa'}-\gamma_{\kappa}-k+1)
\right]. \quad
\label{4.35}
\end{eqnarray}
By performing similar calculations for the second radial integral on the right-hand side of formula in Eq.\ (\ref{4.31}), one has
\begin{eqnarray}
\int_0^{\infty} \textrm{d}r \: r  
\left[
\mu_{n' \kappa'}^{(0)}Q_{n\kappa}^{(0)}(r) S_{n'\kappa'}^{(0)}(r)+P_{n\kappa}^{(0)}(r) T_{n'\kappa'}^{(0)}(r) 
\right]
=\frac{\alpha a_0^2}{4Z}\frac{\mathsf{f}_{n\kappa}\mathsf{g}_{n\kappa}\Gamma(n+2\gamma_{\kappa})N_{n\kappa}}{(|n'|-1)!(N_{n'\kappa'}+\kappa')}  
\sum_{p=0}^{n}\frac{(-)^p}{p!}\frac{\Gamma(\gamma_{\kappa}+\gamma_{\kappa'}+p+2)}{\Gamma(\gamma_{\kappa'}-\gamma_{\kappa}-p-1)}
\nonumber \\
\times  
\frac{\Gamma(|n'|+\gamma_{\kappa'}-\gamma_{\kappa}-p-2)}{(n-p)!\Gamma(p+2\gamma_{\kappa}+1)}  
\left\{
(\mu_{n' \kappa'}^{(0)}+1) 
\left[
(n-p)(N_{n'\kappa'}+\kappa')+(\kappa-N_{n \kappa})(|n'|+\gamma_{\kappa'}-\gamma_{\kappa}-p-2)
\right]\right.
\nonumber \\
\left.
-(\mu_{n' \kappa'}^{(0)}-1)
\left[
(\kappa-N_{n \kappa})(N_{n'\kappa'}+\kappa')+(n-p)(|n'|+\gamma_{\kappa'}-\gamma_{\kappa}-p-2)
\right]
\right\}. 
\label{4.36}
\end{eqnarray}
With the above two results, we are able to start with evaluating the expression for $\mathsf{R}_{\kappa'}$. At first we plug Eqs.\ (\ref{2.12}), (\ref{4.6}), (\ref{4.10}), (\ref{4.35}), and (\ref{4.36}) into Eq.\ (\ref{4.31}). With the help of Eq.\ (\ref{4.11}) and the simple but extremely useful identity
\begin{equation}
\gamma_{\kappa'}^2-\gamma_{\kappa}^2=\kappa'^2-\kappa^2,
\label{4.37}
\end{equation}
we arrive at
\begin{eqnarray}
\mathsf{R}_{\kappa'}&=&\frac{\alpha^2 Z}{N_{n\kappa}^2}\frac{n!\Gamma(n+2\gamma_{\kappa}+1)}{4(N_{n\kappa}-\kappa)}\sum_{k=0}^{n}\sum_{p=0}^{n}
\frac{\mathcal{X}_{\kappa \kappa'}^{n}(k)\mathcal{Y}_{\kappa \kappa'}^{n}(p)}{\Gamma(\gamma_{\kappa'}-\gamma_{\kappa}-k+2)\Gamma(\gamma_{\kappa'}-\gamma_{\kappa}-p-1)}
\nonumber\\
&& \times 
\sum_{n'=-\infty}^{\infty}\frac{\Gamma(|n'|+\gamma_{\kappa'}-\gamma_{\kappa}-k+1)\Gamma(|n'|+\gamma_{\kappa'}-\gamma_{\kappa}-p-2)}{|n'|!\Gamma(|n'|+2\gamma_{\kappa'}+1) (|n'|+\gamma_{\kappa'}-\gamma_{\kappa}-n)} \frac{\kappa'-N_{n' \kappa'}}{N_{n' \kappa'}} 
\nonumber \\
&&\times  
\left[
(n-k)(N_{n'\kappa'}+\kappa')+(\kappa-N_{n \kappa})(|n'|+\gamma_{\kappa'}-\gamma_{\kappa}-k+1)
\right] 
\nonumber \\
&&\times   
\left\{
(|n'|+\gamma_{\kappa'}-\gamma_{\kappa}-n)
\left[
(\kappa-N_{n \kappa})(N_{n'\kappa'}+\kappa') +(n-p)(|n'|+\gamma_{\kappa'}-\gamma_{\kappa}-p-2)
\right]\right.
\nonumber \\
&& \quad 
\left.
-(N_{n'\kappa'}+N_{n \kappa}) 
\left[
(n-p)(N_{n'\kappa'}+\kappa')+(\kappa-N_{n \kappa})(|n'|+\gamma_{\kappa'}-\gamma_{\kappa}-p-2)
\right]
\right\},
\label{4.38}
\end{eqnarray}
where we have defined
\begin{equation}
\mathcal{X}_{\kappa \kappa'}^{n}(k)=\frac{(-)^{k}}{k!(n-k)!} \frac{\Gamma(\gamma_{\kappa}+\gamma_{\kappa'}+k-1)}{\Gamma(k+2\gamma_{\kappa}+1)} \qquad \textrm{and} \qquad \mathcal{Y}_{\kappa \kappa'}^{n}(p)=\frac{(-)^{p}}{p!(n-p)!} \frac{\Gamma(\gamma_{\kappa}+\gamma_{\kappa'}+p+2)}{\Gamma(p+2\gamma_{\kappa}+1)}.
\label{4.39}
\end{equation}
To achieve further simplifications of the formula for $\mathsf{R}_{\kappa'}$, in the infinite series appearing in Eq.\ (\ref{4.38}) we shall collect together terms with the same absolute value of $n'$. After such rearrangements, consequently supported by using Eq.\ (\ref{4.11}) and the identity (\ref{4.37}), we obtain
\begin{eqnarray}
\mathsf{R}_{\kappa'}&=&\frac{\alpha^2 Z}{N_{n\kappa}^2}\frac{n!\Gamma(n+2\gamma_{\kappa}+1)}{2(N_{n\kappa}-\kappa)}\sum_{k=0}^{n}\sum_{p=0}^{n} 
\frac{\mathcal{X}_{\kappa \kappa'}^{n}(k)\mathcal{Y}_{\kappa \kappa'}^{n}(p)}{\Gamma(\gamma_{\kappa'}-\gamma_{\kappa}-k+2)\Gamma(\gamma_{\kappa'}-\gamma_{\kappa}-p-1)} {}
\nonumber \\
&& \times 
\sum_{n'=0}^{\infty}\frac{\Gamma(n'+\gamma_{\kappa'}-\gamma_{\kappa}-k+1)\Gamma(n'+\gamma_{\kappa'}-\gamma_{\kappa}-p-2)}{n'!\Gamma(n'+2\gamma_{\kappa'}+1)(n'+\gamma_{\kappa'}-\gamma_{\kappa}-n)}
\nonumber \\
&& \times 
\left\{
(N_{n\kappa}-\kappa)^2(N_{n\kappa}-\kappa')(n'+\gamma_{\kappa'}-\gamma_{\kappa}-k+1)(n'+\gamma_{\kappa'}-\gamma_{\kappa}-p-2)
\right.
\nonumber \\	
&& \quad 
+n'(n'+2\gamma_{\kappa'}) 
\left[
(n-k)(n-p)(2\kappa+\kappa'-N_{n\kappa})+(n+p-2k)(N_{n\kappa}-\kappa)
\right]
\nonumber \\
&&\quad
+(n-p)(N_{n\kappa}-\kappa)(n'+\gamma_{\kappa'}-\gamma_{\kappa}-n)(n'+\gamma_{\kappa'}-\gamma_{\kappa}-k+1)(n'+\gamma_{\kappa'}-\gamma_{\kappa}-p-2)
\nonumber	\\
&&\quad 
\left.
-(n-p)(N_{n\kappa}-\kappa)(n'+\gamma_{\kappa'}-\gamma_{\kappa}-n)n'(n'+2\gamma_{\kappa'})	
\right\}.
\label{4.40}
\end{eqnarray}
Utilizing the recurrence relation $\Gamma(\zeta+1)=\zeta\Gamma(\zeta)$, one can easily show that the sum of the two series $\sum_{n'=0}^{\infty}(...)$, formed by the last two components between the curly braces, equals zero. In turn, employing the aforementioned relation satisfied by the gamma functions to the series produced with the use of the remaining two terms, one may recognize therein the hypergeometric functions ${}_3F_2$ with the unit arguments. According to the definition of these special functions, which is \cite{Bail35,Slat66}
\begin{eqnarray}
 {}_3F_2 
\left(
\begin{array}{c} 
a_1, a_2, a_3\\
b_1, b_2
\end{array}
;1 
\right)
=\frac{\Gamma(b_1)\Gamma(b_2)}{\Gamma(a_1) \Gamma(a_2)\Gamma(a_3)} 
\sum_{n=0}^{\infty} \frac{\Gamma(a_1+n)\Gamma(a_2+n)\Gamma(a_3+n)}{n!\Gamma(b_1+n)\Gamma(b_2+n)}
\quad 
\left[\Real\left(b_1+b_2-a_1-a_2-a_3 \right)>0\right], \qquad 
\label{4.41} 
\end{eqnarray}
we rewrite Eq.\ (\ref{4.40}) in the following form: 
\begin{eqnarray}
\mathsf{R}_{\kappa'}&=&\frac{\alpha^2 Z}{2N_{n\kappa}^2}\frac{n!\Gamma(n+2\gamma_{\kappa}+1)}{(N_{n\kappa}-\kappa)\Gamma(2\gamma_{\kappa'}+1)}\sum_{k=0}^{n}\sum_{p=0}^{n} 
\mathcal{X}_{\kappa \kappa'}^{n}(k)\mathcal{Y}_{\kappa \kappa'}^{n}(p)
\nonumber \\
&& \times 
\Bigg\{
\frac{(N_{n\kappa}-\kappa')(N_{n\kappa}-\kappa)^2}{\gamma_{\kappa'}-\gamma_{\kappa}-n} 
{}_3F_2 
\left(
\begin{array}{c} 
\gamma_{\kappa'}-\gamma_{\kappa}-k+2,\: 
\gamma_{\kappa'}-\gamma_{\kappa}-p-1,\: 
\gamma_{\kappa'}-\gamma_{\kappa}-n\\
\gamma_{\kappa'}-\gamma_{\kappa}-n+1,\:
2\gamma_{\kappa'}+1
\end{array}
;1 
\right)
\nonumber \\
&& \quad 
+\frac{(n-k)(n-p)(2\kappa+\kappa'-N_{n\kappa})+(n+p-2k)(N_{n\kappa}-\kappa)}{\gamma_{\kappa'}-\gamma_{\kappa}-n+1}
\nonumber \\
&&\quad 
\times {}_3F_2 
\left(
\begin{array}{c} 
\gamma_{\kappa'}-\gamma_{\kappa}-k+2,\: 
\gamma_{\kappa'}-\gamma_{\kappa}-p-1,\: 
\gamma_{\kappa'}-\gamma_{\kappa}-n+1 \\
\gamma_{\kappa'}-\gamma_{\kappa}-n+2,\:
2\gamma_{\kappa'}+1
\end{array}
;1 
\right)
\Bigg\}.
\label{4.42}
\end{eqnarray}
To eliminate the first ${}_3F_{2}$ function, we have to invoke the recurrence formula 
\begin{eqnarray}
{}_3F_2 
\left( 
\begin{array}{c} 
a_1, a_2, a_3-1\\
a_3,b
\end{array}
;1 
\right)
=-\frac{(a_1-a_3)(a_2-a_3)}{a_3(b-a_3)}  
{}_3F_2 
\left(
\begin{array}{c} 
a_1, a_2, a_3\\
a_3+1,b
\end{array}
;1 
\right)
+\frac{\Gamma(b)\Gamma(b-a_1-a_2+1)}{(b-a_3)\Gamma(b-a_1)\Gamma(b-a_2)}
\nonumber \\ 
\left[\Real(b-a_1-a_2)>-1\right]. 
\label{4.43}
\end{eqnarray}
Proceeding in that way, after some algebra,  Eq.\ (\ref{4.42}) becomes
\begin{eqnarray}
\mathsf{R}_{\kappa'}&=&\frac{\alpha^2 Z}{2N_{n\kappa}^2}\frac{N_{n \kappa}-\kappa}{N_{n \kappa}+\kappa'}
\Bigg\{
\mathcal{F}_{\kappa}^{n}(1)-\frac{n!\Gamma(n+2\gamma_{\kappa}+1)}{(N_{n\kappa}-\kappa)^2(\gamma_{\kappa'}-\gamma_{\kappa}-n+1)\Gamma(2\gamma_{\kappa'}+1)}
\nonumber \\
&& 
\times
\sum_{k=0}^n\sum_{p=0}^n \widetilde{\mathcal{X}}_{\kappa \kappa'}^{n}(k)\widetilde{\mathcal{Y}}_{\kappa \kappa'}^{n}(p) 
{}_3F_2 
\left(
\begin{array}{c} 
\gamma_{\kappa'}-\gamma_{\kappa}-k+2,\: 
\gamma_{\kappa'}-\gamma_{\kappa}-p-1,\: 
\gamma_{\kappa'}-\gamma_{\kappa}-n+1 \\
\gamma_{\kappa'}-\gamma_{\kappa}-n+2,\:
2\gamma_{\kappa'}+1
\end{array}
;1 
\right)
\Bigg\},
\label{4.44}
\end{eqnarray}
where 
\begin{equation}
\widetilde{\mathcal{X}}_{\kappa \kappa'}^{n}(k)=
\left
[(N_{n \kappa}-\kappa)-(n-k)(\kappa+\kappa')
\right]
\mathcal{X}_{\kappa \kappa'}^{n}(k),
\label{4.45}
\end{equation}
\begin{equation}
\widetilde{\mathcal{Y}}_{\kappa \kappa'}^{n}(p)=
\left
[2(N_{n \kappa}-\kappa)+(n-p)(\kappa+\kappa')
\right]
\mathcal{Y}_{\kappa \kappa'}^{n}(p)
\label{4.46}
\end{equation}
and
\begin{equation}
\mathcal{F}_{\kappa}^{n}(1)= n!\Gamma(n+2\gamma_{\kappa}+1)
\sum_{k=0}^n\sum_{p=0}^n\frac{(-)^{k+p+1}\Gamma(2\gamma_{\kappa}+k+p+1)}{k!p!(n-p)!(n-k)!\Gamma(k+2\gamma_{\kappa}+1)\Gamma(p+2\gamma_{\kappa}+1)}.
\label{4.47}
\end{equation}
Basing on the analysis of such kind of expressions as the one in the last equation, carried out by us in the appendix to Ref.\ \cite{Stef15}, particularly, using Eq.\ (A.11) from that article, one may easily prove that $\mathcal{F}_{\kappa}^{n}(1)=-1$. Utilizing this fact in Eq.\ (\ref{4.44}) and transforming the hypergeometric function appearing therein with the aid of the following relation (see Eq.\ (3.2.1) in Ref.\ \cite{Bail35} or Eq.\ (7.4.4.1) in Ref.\ \cite{Prud03}):
  \begin{eqnarray}
 {}_3F_2 \left(\begin{array}{c} a_1, a_2, a_3 \\ b_1, b_2 \end{array};1\right)=\frac{\Gamma(b_1)\Gamma(b_1+b_2-a_1-a_2-a_3)}{\Gamma(b_1+b_2-a_1-a_2)\Gamma(b_1-a_3)}{}_3F_2 \left( \begin{array}{c} b_2-a_1,b_2-a_2, a_3 \\b_1+b_2-a_1-a_2, b_2 \end{array};1\right) {}
 \nonumber
 \\ \left[\textrm{Re}(b_1+b_2-a_1-a_2-a_3)>0, \ \textrm{Re}(b_1-a_3)>0 \right],
 \label{4.48}
 \end{eqnarray}
we obtain
\begin{eqnarray}
\mathsf{R}_{\kappa'}=\frac{\alpha^2 Z}{2N_{n\kappa}^2}\frac{\kappa-N_{n \kappa}}{\kappa'+N_{n \kappa}}
\Bigg\{
1+\frac{n!\Gamma(n+2\gamma_{\kappa}+1)}{(N_{n\kappa}-\kappa)^2(\gamma_{\kappa'}-\gamma_{\kappa}-n+1)\Gamma(n+\gamma_{\kappa}+\gamma_{\kappa'})}
\sum_{k=0}^n\sum_{p=0}^n \widetilde{\mathcal{X}}_{\kappa \kappa'}^{n}(k)\widetilde{\mathcal{Y}}_{\kappa \kappa'}^{n}(p) {}
\nonumber \\
 \times \frac{\Gamma(2\gamma_{\kappa}+k+p+1)}{\Gamma(\gamma_{\kappa}+\gamma_{\kappa'}+k+p-n+2)} 
{}_3F_2 
\left(
\begin{array}{c} 
k-n,\quad 
p-n+3,\quad  
\gamma_{\kappa'}-\gamma_{\kappa}-n+1 \\
\gamma_{\kappa}+\gamma_{\kappa'}+k+p-n+2, \  
\gamma_{\kappa'}-\gamma_{\kappa}-n+2
\end{array}
;1 
\right)
\Bigg\}.
\label{4.49}
\end{eqnarray}
Notice, that in view of the summation limits, the value of $k-n$ is a nonpositive integer. Hence, the above ${}_3F_2$ function may be represented as a finite sum of ratios of Euler's gamma functions. It follows that for any value $n$ of the radial quantum number, the formula for $\mathsf{R}_{\kappa'}$ appearing in Eq.\ (\ref{4.49}), may be readily programmed in order to obtain numerical values of $\sigma$. Nevertheless, in the following part of this paper, we provide further analytical calculations, after which the current expression for $\mathsf{R}_{\kappa'}$, and consequently for the total magnetic shielding constant $\sigma$, will be simplified to an elementary form.  

Let us focus our attention on the ${}_3F_2$ function in the above equation. It will be temporary eliminated with the aid of the formula
\begin{equation}
{}_3F_2 
\left(
\begin{array}{c}
 a_1, a_2, a_3 
\\ a_3+1, b 
\end{array}
;1
\right)
=a_3 \int_{0}^{1} \textrm{d}x \: x^{a_3-1} 
{}_2F_1 
\left(
\begin{array}{c}
 a_1, a_2 
\\ b
\end{array}
;x
\right),
\label{4.50}
\end{equation}
in which ${}_2F_1$ is the hypergeometric Gauss' function, defined as
	\begin{equation}
	{}_2F_1 
	\left(
	\begin{array}{c} 
	a_1,a_2 
	\\ b 
	\end{array}
	;x \right)
	=\frac{\Gamma(b)}{\Gamma(a_1)\Gamma(a_2)} \sum_{m=0}^{\infty} \frac{\Gamma(m+a_1)\Gamma(m+a_2)}{\Gamma(m+b)} \frac{x^m}{m!} \quad \quad \left(|x| \leqslant 1\right).
	 \label{4.51}
	\end{equation}
After these two steps, i.e., utilizing relation (\ref{4.50}) to the ${}_3F_2$ function from Eq.\ (\ref{4.49}) and rewriting the resulting ${}_2F_1$ function with the help of the definition in Eq.\ (\ref{4.51}), we arrive at
\begin{eqnarray}
\mathsf{R}_{\kappa'}=\frac{\alpha^2 Z}{2N_{n\kappa}^2}\frac{\kappa-N_{n \kappa}}{\kappa'+N_{n \kappa}}
\left\{
1+\frac{n!\Gamma(n+2\gamma_{\kappa}+1)}{(N_{n\kappa}-\kappa)^2\Gamma(n+\gamma_{\kappa}+\gamma_{\kappa'})}
\sum_{k=0}^n\sum_{p=0}^n \frac{\widetilde{\mathcal{X}}_{\kappa \kappa'}^{n}(k)\widetilde{\mathcal{Y}}_{\kappa \kappa'}^{n}(p)\Gamma(2\gamma_{\kappa}+k+p+1)}{\Gamma(k-n)\Gamma(p-n+3)} \right.
\nonumber \\ 
\times \left.
\sum_{m=0}^{\infty} \frac{\Gamma(m+k-n)\Gamma(m+p-n+3)}{m!\Gamma(m+\gamma_{\kappa}+\gamma_{\kappa'}+k+p-n+2)}
\int_{0}^{1} \textrm{d}x \: x^{m+\gamma_{\kappa'}-\gamma_{\kappa}-n}
\right\}.
\label{4.52}
\end{eqnarray}
The integral, which has emerged here, equals $\left(m+\gamma_{\kappa'}-\gamma_{\kappa}-n+1\right)^{-1}$. On putting this result into the last equation and employing the first formula from Eq.\ (\ref{4.39}), the above expression becomes
\begin{eqnarray}
\mathsf{R}_{\kappa'}&=&\frac{\alpha^2 Z}{2N_{n\kappa}^2}\frac{\kappa-N_{n \kappa}}{\kappa'+N_{n \kappa}}
\left\{
1+\frac{n!\Gamma(n+2\gamma_{\kappa}+1)}{(N_{n\kappa}-\kappa)^2\Gamma(n+\gamma_{\kappa}+\gamma_{\kappa'})} \sum_{p=0}^n \frac{\widetilde{\mathcal{Y}}_{\kappa \kappa'}^{n}(p)}{\Gamma(p-n+3)} \sum_{m=0}^{\infty} \frac{\Gamma(m+p-n+3)}{m!(m+\gamma_{\kappa'}-\gamma_{\kappa}-n+1)} 
\right.
\nonumber \\
&& \times 
\left.
\sum_{k=0}^{\infty} \frac{(-)^k}{k!} \frac{\Gamma(k+\gamma_{\kappa}+\gamma_{\kappa'}-1)\Gamma(k+2\gamma_{\kappa}+p+1)\Gamma(m+k-n)}{\Gamma(k+2\gamma_{\kappa}+1)\Gamma(k+\gamma_{\kappa}+\gamma_{\kappa'}+m+p-n+2)} \frac{N_{n\kappa}-\kappa-(n-k)(\kappa+\kappa')}{\displaystyle{\lim_{\nu \to n}{\Gamma(\nu-k+1)\Gamma(k-\nu)}}}
\right\},
\label{4.53}
\end{eqnarray}
where we have used the fact, that the components with $k  \in [n+1, \infty)$ equal zero. If the reflection relation obeyed by the gamma functions
\begin{equation}
\Gamma(\zeta)\Gamma(1-\zeta)=\frac{\pi}{\sin{\pi \zeta}}
\label{4.54}
\end{equation}
is applied to the expression appearing under the limit procedure, the current formula for $\mathsf{R}_{\kappa'}$ may be written as follows:
\begin{eqnarray}
\mathsf{R}_{\kappa'}=\frac{\alpha^2 Z}{2N_{n\kappa}^2}\frac{\kappa-N_{n \kappa}}{\kappa'+N_{n \kappa}}
\left\{
1-\frac{n!\Gamma(n+2\gamma_{\kappa}+1)}{(N_{n\kappa}-\kappa)^2\Gamma(n+\gamma_{\kappa}+\gamma_{\kappa'})} 
\lim_{\nu \to n}{\frac{\sin(\pi \nu)}{\pi}}
\sum_{p=0}^n \frac{\widetilde{\mathcal{Y}}_{\kappa \kappa'}^{n}(p)}{\Gamma(p-n+3)} 
\right.
\nonumber \\
 \times 
\left. 
\sum_{m=0}^{\infty} \frac{\Gamma(m+p-n+3)}{m!(m+\gamma_{\kappa'}-\gamma_{\kappa}-n+1)}
\Big[(\kappa+\kappa')\textrm{S}_{mp}^{(1)}+[N_{n\kappa}-\kappa-n(\kappa'+\kappa)]\textrm{S}_{mp}^{(2)}\Big]
\right\},
\label{4.55}
\end{eqnarray}
with
\begin{equation}
\textrm{S}_{mp}^{(1)}=\sum_{k=0}^{\infty} \frac{\Gamma(k+\gamma_{\kappa}+\gamma_{\kappa'})\Gamma(k+2\gamma_{\kappa}+p+2)\Gamma(k+m-n+1)}{k!\Gamma(k+2\gamma_{\kappa}+2)\Gamma(k+\gamma_{\kappa}+\gamma_{\kappa'}+m+p-n+3)} 
\label{4.56} 
\end{equation}
and
\begin{equation}
\textrm{S}_{mp}^{(2)}=\sum_{k=0}^{\infty} \frac{\Gamma(k+\gamma_{\kappa}+\gamma_{\kappa'}-1)\Gamma(k+2\gamma_{\kappa}+p+1)\Gamma(k+m-n)}{k!\Gamma(k+2\gamma_{\kappa}+1)\Gamma(k+\gamma_{\kappa}+\gamma_{\kappa'}+m+p-n+2)}. 
\label{4.57} 
\end{equation}
In view of the definition given in Eq.\ (\ref{4.41}), we rewrite $\textrm{S}_{mp}^{(1)}$ in the form of
\begin{equation}
\textrm{S}_{mp}^{(1)}=\frac{\Gamma(\gamma_{\kappa}+\gamma_{\kappa'})\Gamma(2\gamma_{\kappa}+p+2)\Gamma(m-n+1)}{\Gamma(2\gamma_{\kappa}+2)\Gamma(m+\gamma_{\kappa}+\gamma_{\kappa'}+p-n+3)} {}_3F_{2}
\left(
\begin{array}{c}
\gamma_{\kappa}+\gamma_{\kappa'}, \: 
2\gamma_{\kappa}+p+2, \: 
m-n+1 \\
2\gamma_{\kappa}+2, \: 
m+\gamma_{\kappa}+\gamma_{\kappa'}+p-n+3
\end{array}
;1
\right).
\label{4.58}
\end{equation}
In the next step, we shall transform the ${}_3F_{2}$ function with the aid of formula in Eq.\ (\ref{4.48}), yielding
\begin{equation}
\textrm{S}_{mp}^{(1)}=\frac{\Gamma(\gamma_{\kappa}+\gamma_{\kappa'})\Gamma(2\gamma_{\kappa}+p+2)\Gamma(m-n+1)}{\Gamma(\gamma_{\kappa}+\gamma_{\kappa'}+p+2)\Gamma(2\gamma_{\kappa}+2)\Gamma(m-n+3)} {}_3F_{2}
\left(
\begin{array}{c}
-p, \: 
\gamma_{\kappa}-\gamma_{\kappa'}+2, \: 
m-n+1 \\
2\gamma_{\kappa}+2, \ 
m-n+3
\end{array}
;1
\right).
\label{4.59}
\end{equation}
Now we can use the relation (see Eq.\ (7.4.4.81) in Ref.\ \cite{Prud03})
\begin{equation}
{}_3F_2 
\left(
\begin{array}{c} 
-l, \: a_1, \;  a_2 \\ 
b_1, \: b_2 
\end{array} 
;1 \right) 
=\frac{(a_1)_l (b_1+b_2-a_1-a_2)_l}{(b_1)_l (b_2)_l}  
{}_3F_2 
\left(
\begin{array}{c} 
-l,\: b_1-a_1,\: b_2-a_1\\ 
1-a_1-l, \: b_1+b_2-a_1-a_2 
\end{array} 
;1 
\right) \qquad \left(l \in \mathbb{N}\right),
\label{4.60}
\end{equation}
in which $(a)_{l}=\Gamma(a+l)/\Gamma(a)$ is the Pochhammer's symbol. By doing so, we get
\begin{equation}
\textrm{S}_{mp}^{(1)}=\frac{\Gamma(\gamma_{\kappa}+\gamma_{\kappa'})\Gamma(\gamma_{\kappa}-\gamma_{\kappa'}+p+2)\Gamma(m-n+1)}{\Gamma(\gamma_{\kappa}+\gamma_{\kappa'}+2)\Gamma(\gamma_{\kappa}-\gamma_{\kappa'}+2)\Gamma(p+m-n+3)} {}_3F_{2}
\left(
\begin{array}{c}
-p, \: 
\gamma_{\kappa}+\gamma_{\kappa'}, \: 
\gamma_{\kappa'}-\gamma_{\kappa}+m-n+1 \\
\gamma_{\kappa'}-\gamma_{\kappa}-p-1, \ 
\gamma_{\kappa}+\gamma_{\kappa'}+2
\end{array}
;1
\right).
\label{4.61}
\end{equation}
Then, taking advantage of the formula [see Eq.\ (7.4.4.85) in Ref.\ \cite{Prud03}]
\begin{equation}
{}_3F_2 
\left(
\begin{array}{c} 
-l, \: a_1, \: a_2 \\ 
b_1, \: b_2 
\end{array} 
;1 
\right) 
=\frac{(b_1-a_1)_l}{ (b_1)_l} {}_3F_2 
\left(
\begin{array}{c} 
-l,\: a_1, \: b_2-a_2 \\ 
b_2, \: a_1-b_1-l+1 
\end{array} 
;1 
\right) \qquad \left(l \in \mathbb{N}\right),
\label{4.62}
\end{equation}
one has
\begin{equation}
\textrm{S}_{mp}^{(1)}=\frac{\Gamma(\gamma_{\kappa}+\gamma_{\kappa'})\Gamma(\gamma_{\kappa}-\gamma_{\kappa'}+p+2)\Gamma(p+2)\Gamma(m-n+1)}{\Gamma(\gamma_{\kappa}-\gamma_{\kappa'}+2)\Gamma(\gamma_{\kappa}+\gamma_{\kappa'}+p+2)\Gamma(p+m-n+3)} {}_3F_{2}
\left(
\begin{array}{c}
-p, \: 
\gamma_{\kappa}+\gamma_{\kappa'}, \: 
n-m-p-2 \\
-p-1, \: 
\gamma_{\kappa'}-\gamma_{\kappa}-p-1
\end{array}
;1
\right).
\label{4.63}
\end{equation}
In this way we arrive at the situation, when one of the upper parameters of the hypergeometric function is increased by one with respect to the one of its lower parameters. Such ${}_3F_{2}$ functions may be easily cast to an elementary form. To this end, in our case, we shall employ the identity
\begin{equation}
{}_3F_2 
\left( 
\begin{array}{c} 
a_1, \; a_2, \; a_3+1 \\
a_3, \: b
\end{array}
;1
\right)
=\frac{\Gamma(b)\Gamma(b-a_1-a_2)}{\Gamma(b-a_1)\Gamma(b-a_2)}\left[1+ \frac{a_1 a_2}{a_3(b-a_1-a_2-1)}\right]
\label{4.64}
\end{equation}
and obtain
\begin{eqnarray}
\textrm{S}_{mp}^{(1)}&=&\frac{\Gamma(\gamma_{\kappa}+\gamma_{\kappa'})\Gamma(\gamma_{\kappa'}-\gamma_{\kappa}-p-1)\Gamma(\gamma_{\kappa}-\gamma_{\kappa'}+p+2)\Gamma(p+1)\Gamma(m-n+1)\Gamma(-2\gamma_{\kappa}+m-n)}{\Gamma(\gamma_{\kappa}-\gamma_{\kappa'}+2)\Gamma(\gamma_{\kappa}+\gamma_{\kappa'}+p+2)\Gamma(-2\gamma_{\kappa}-p-1)\Gamma(p+m-n+3)\Gamma(\gamma_{\kappa'}-\gamma_{\kappa}+m-n+1)} 
\nonumber \\
&&\times
\left[
(p+1)(-2\gamma_{\kappa}+m-n)+(\gamma_{\kappa}+\gamma_{\kappa'})(p+m-n+2)
\right].
\label{4.65}
\end{eqnarray}
Proceeding in a similar way, one can show that
\begin{eqnarray}
\textrm{S}_{mp}^{(2)}&=&\frac{\Gamma(\gamma_{\kappa}+\gamma_{\kappa'}-1)\Gamma(\gamma_{\kappa'}-\gamma_{\kappa}-p-1)\Gamma(\gamma_{\kappa}-\gamma_{\kappa'}+p+2)\Gamma(p+1)\Gamma(m-n)\Gamma(-2\gamma_{\kappa}+m-n)}{\Gamma(\gamma_{\kappa}-\gamma_{\kappa'}+2)\Gamma(\gamma_{\kappa}+\gamma_{\kappa'}+p+2)\Gamma(-2\gamma_{\kappa}-p)\Gamma(p+m-n+3)\Gamma(\gamma_{\kappa'}-\gamma_{\kappa}+m-n+1)} 
\nonumber \\
&&\times
\left\{
(p+1)(p+2)(-2\gamma_{\kappa}+m-n)(-2\gamma_{\kappa}+m-n+1) \right.
\nonumber \\
&& \quad \left.
+2(\gamma_{\kappa}+\gamma_{\kappa'}-1)(p+1)(p+m-n+2)(-2\gamma_{\kappa}+m-n) \right.
\nonumber \\
&& \quad \left.
+(\gamma_{\kappa}+\gamma_{\kappa'})(\gamma_{\kappa}+\gamma_{\kappa'}-1)(p+m-n+2)(p+m-n+1)
\right\}.
\label{4.66}
\end{eqnarray}
If the last two results are inserted into Eq.\ (\ref{4.55}), after much labor, the expression for $\mathsf{R}_{\kappa'}$ becomes
\begin{eqnarray}
\mathsf{R}_{\kappa'}&=&\frac{\alpha^2 Z}{2N_{n\kappa}^2}\frac{\kappa-N_{n \kappa}}{\kappa'+N_{n \kappa}}
\left\{
1-\frac{n!\Gamma(n+2\gamma_{\kappa}+1)}{(N_{n\kappa}-\kappa)^2\Gamma(n+\gamma_{\kappa}+\gamma_{\kappa'})} \frac{\Gamma(\gamma_{\kappa}+\gamma_{\kappa'}-1)}{\Gamma(\gamma_{\kappa}-\gamma_{\kappa'}+2)}
\lim_{\nu \to n}{\frac{\sin(\pi \nu)}{\pi}}
\sum_{p=0}^n  \frac{\widetilde{\mathcal{Y}}_{\kappa \kappa'}^{n}(p) p!}{(p-n+2)!} \right. 
\nonumber \\
&&\times \left.
\frac{\Gamma(\gamma_{\kappa'}-\gamma_{\kappa}-p-1)\Gamma(\gamma_{\kappa}-\gamma_{\kappa'}+p+2)}{\Gamma(-2\gamma_{\kappa}-p)\Gamma(\gamma_{\kappa}+\gamma_{\kappa'}+p+2)}
\left[
\textrm{C}_{p}^{(\mathsf{I})} \sum_{m=0}^{\infty} \frac{\Gamma(m-n+1)\Gamma(m-2\gamma_{\kappa}-n+1)}{m!\Gamma(m+\gamma_{\kappa'}-\gamma_{\kappa}-n+2)} \right. \right.
\nonumber \\
&& \left. \left. \quad  + \textrm{C}_{p}^{(\mathsf{II})} \sum_{m=0}^{\infty} \frac{\Gamma(m-n+1)\Gamma(m-2\gamma_{\kappa}-n)}{m!\Gamma(m+\gamma_{\kappa'}-\gamma_{\kappa}-n+2)}
+\textrm{C}_{p}^{(\mathsf{III})} \sum_{m=0}^{\infty} \frac{\Gamma(m-n)\Gamma(m-2\gamma_{\kappa}-n)}{m!\Gamma(m+\gamma_{\kappa'}-\gamma_{\kappa}-n+2)}
\right]
\right\},
\label{4.67}
\end{eqnarray}
where
\begin{equation}
\textrm{C}_{p}^{(\mathsf{I})}=(\gamma_{\kappa}+\gamma_{\kappa'}+p+1)
\left\{
(\gamma_{\kappa}+\gamma_{\kappa'}+p) \left[N_{n\kappa}-\kappa-n(\kappa+\kappa')\right]-(\kappa+\kappa')(\gamma_{\kappa}+\gamma_{\kappa'}-1)(p+2\gamma_{\kappa}+1)
\right\},
\label{4.68}
\end{equation}
\begin{eqnarray}
\textrm{C}_{p}^{(\mathsf{II})}&=&(\gamma_{\kappa}+\gamma_{\kappa'}+p+1)
\left[
N_{n\kappa}-\kappa-n(\kappa+\kappa')
\right]
\left[
(\gamma_{\kappa'}-\gamma_{\kappa})(p+1)+(\gamma_{\kappa'}+\gamma_{\kappa}-1)(p+2\gamma_{\kappa}+2)
\right]
\nonumber \\
&& -(\kappa+\kappa')(\gamma_{\kappa}+\gamma_{\kappa'})(\gamma_{\kappa}+\gamma_{\kappa'}-1)(p+2\gamma_{\kappa}+1)(p+2\gamma_{\kappa}+2),
\label{4.69}
\end{eqnarray}
\begin{equation}
\textrm{C}_{p}^{(\mathsf{III})}=(p+1)(p+2)(\gamma_{\kappa'}-\gamma_{\kappa}-1)(\gamma_{\kappa'}-\gamma_{\kappa})
\left[
N_{n\kappa}-\kappa-n(\kappa+\kappa')
\right].
\label{4.70}
\end{equation}
In view of the definition in Eq.\ (\ref{4.51}), the components between the square braces in Eq.\ (\ref{4.67}) may be written in terms of ${}_2F_{1}$ functions with the unit arguments, and consequently
\begin{eqnarray}
\mathsf{R}_{\kappa'}&=&\frac{\alpha^2 Z}{2N_{n\kappa}^2}\frac{\kappa-N_{n \kappa}}{\kappa'+N_{n \kappa}}
\Bigg\{
1+\frac{\Gamma(n+2\gamma_{\kappa}+1)\Gamma(-2\gamma_{\kappa}-n)}{(N_{n\kappa}-\kappa)^2\Gamma(n+\gamma_{\kappa}+\gamma_{\kappa'})\Gamma(\gamma_{\kappa'}-\gamma_{\kappa}-n+2)} \frac{\Gamma(\gamma_{\kappa}+\gamma_{\kappa'}-1)}{\Gamma(\gamma_{\kappa}-\gamma_{\kappa'}+2)}
\sum_{p=0}^n  \frac{\widetilde{\mathcal{Y}}_{\kappa \kappa'}^{n}(p) p!}{(p-n+2)!}  
\nonumber \\
&&\times \left.
\frac{\Gamma(\gamma_{\kappa'}-\gamma_{\kappa}-p-1)\Gamma(\gamma_{\kappa}-\gamma_{\kappa'}+p+2)}{\Gamma(-2\gamma_{\kappa}-p)\Gamma(\gamma_{\kappa}+\gamma_{\kappa'}+p+2)}
\left[
  n(n+2\gamma_{\kappa})  \textrm{C}_{p}^{(\mathsf{I})}   {}_2F_{1}\left(\begin{array}{c} -n+1, \: -2\gamma_{\kappa}-n \\ \gamma_{\kappa'}-\gamma_{\kappa}-n+2 \end{array};1 \right)
 \right. \right.
\nonumber \\
&& \left. \quad  -n  \textrm{C}_{p}^{(\mathsf{II})}  {}_2F_{1}\left(\begin{array}{c} -n+1, \: -2\gamma_{\kappa}-n+1 \\ \gamma_{\kappa'}-\gamma_{\kappa}-n+2 \end{array};1 \right) 
+ \textrm{C}_{p}^{(\mathsf{III})}  {}_2F_{1}\left(\begin{array}{c} -n, \: -2\gamma_{\kappa}-n \\ \gamma_{\kappa'}-\gamma_{\kappa}-n+2 \end{array};1 \right) 
\right]
\Bigg\}.
\label{4.71}
\end{eqnarray}
Employing the Gauss' identity
\begin{equation}
 {}_2F_1 \left( \begin{array}{c} 
a_1,a_2\\
b
\end{array}
;1 
\right)
=\frac{\Gamma(b)\Gamma(b-a_1-a_2)}{\Gamma(b-a_1)\Gamma(b-a_2)} \qquad \qquad \left[\textrm{Re} (b-a_1-a_2)>0 \right],
 \label{4.72}
\end{equation}
taking advantage of Eqs.\ (\ref{4.39}) and (\ref{4.46}), after some rearrangements, involving the recurrence relation $\Gamma(\zeta+1)=\zeta\Gamma(\zeta)$ and the identities (\ref{4.37}) and (\ref{4.54}), we arrive at
\begin{eqnarray}
\mathsf{R}_{\kappa'}&=&\frac{\alpha^2 Z}{2N_{n\kappa}^2}\frac{\kappa-N_{n \kappa}}{\kappa'+N_{n \kappa}}
\Big\{1+\frac{(-)^{n+1}}{(N_{n\kappa}-\kappa)^2(\kappa'^2-\kappa^2)(\kappa^2-\kappa'^2+2\gamma_{\kappa}+1)(\kappa'^2-\kappa^2+2\gamma_{\kappa}-1)}  
\nonumber \\
&&\times 
\sum_{p=n-2}^{n} \frac{(-)^p [2(N_{n\kappa}-\kappa)+(n-p)(\kappa+\kappa')]}{(n-p)!(p-n+2)!} 
\left[
   n(n+2\gamma_{\kappa})(\gamma_{\kappa'}+\gamma_{\kappa}+1)(\gamma_{\kappa'}-\gamma_{\kappa}+1)\textrm{C}_{p}^{(\mathsf{I})}  
 \right. 
\nonumber \\
&& \left. \quad 
-n(\gamma_{\kappa'}+\gamma_{\kappa}+n)(\gamma_{\kappa'}-\gamma_{\kappa}+1)  \textrm{C}_{p}^{(\mathsf{II})}      
+(\gamma_{\kappa'}+\gamma_{\kappa}+n)(\gamma_{\kappa'}+\gamma_{\kappa}+n+1) \textrm{C}_{p}^{(\mathsf{III})} \right] 
\Big\}, 
\label{4.73}
\end{eqnarray}
where it has been marked, that the only nonzero components of the above sum are those with $p=n, n-1, n-2$. By plugging Eqs.\ (\ref{4.68})--(\ref{4.70}) into the above formula and adding all the aforementioned nonvanishing components, after much labor, we obtain finally
\begin{equation}
\mathsf{R}_{\kappa'}=\frac{\alpha^2 Z}{N_{n\kappa}^2}\frac{4(n+\gamma_{\kappa})(N_{n\kappa}-\kappa)(N_{n\kappa}-\kappa')-[2n(n+2\gamma_{\kappa})-\kappa(\kappa+\kappa')\Delta_{n\kappa}^{(-)}]\Delta_{n\kappa}^{(+)}}{(\kappa'^2-\kappa^2)(\kappa^2-\kappa'^2+2\gamma_{\kappa}+1)(\kappa'^2-\kappa^2+2\gamma_{\kappa}-1)},
\label{4.74}
\end{equation}
with $\Delta_{n\kappa}^{(\pm)}=2n+2\gamma_{\kappa}-1\pm\kappa^2\mp\kappa'^2$. On putting this result into Eq.\ (\ref{4.30}), with some algebra, we eventually find that the sum of the two remaining components of the magnetic shielding constant equals
\begin{equation}
\sigma_{-\kappa-1}+\sigma_{-\kappa+1}=\frac{\alpha^2 Z}{N_{n\kappa}^2(4\kappa^2-1)} 
\left[ 
\kappa^2-\frac{\eta_{n\kappa}^{(+)}}{4}-\frac{\eta_{n\kappa}^{(-)}}{4}
+\frac{\mu^2}{4\kappa^2-1}
\left(
\frac{2\kappa+1}{2\kappa-1}\eta_{n\kappa}^{(+)}+\frac{2\kappa-1}{2\kappa+1}\eta_{n\kappa}^{(-)}-4\kappa^2\frac{4\kappa^2+3}{4\kappa^2-1}
\right)
\right],
\label{4.75}
\end{equation}
where we define 
\begin{equation}
\eta_{n\kappa}^{(\pm)}=\frac{(2\kappa \pm 1)N_{n\kappa}}{n+\gamma_{\kappa}\pm N_{n\kappa}}.
\label{4.76}
\end{equation}
In analogy to the test performed for the constituent $\sigma_{\kappa}$, we will check the above expression for the atomic ground state. In this particular case, i.e., for $n=0$, $\kappa=-1$, $\mu=\pm 1/2$, Eq.\ (\ref{4.75}) becomes
\begin{equation}
\sigma_{+2}=\frac{2 \alpha^2 Z}{27} \frac{\gamma_{1}+2}{\gamma_{1}+1}.
\label{4.77}
\end{equation}
The formula for $\sigma_{+2}$ presented by Cheng \emph{et al.} in Ref.\ \cite{Chen09}, after some algebraic simplification provided in Ref.\ \cite{Szmy11}, takes the identical form with the one given in Eq.\ (\ref{4.77}). 

After this digression, concerning the part of the formula for the total magnetic shielding constant of the relativistic one-electron atom, we can write the final expression for the latter. To this end, we combine Eqs.\ (\ref{4.28}) and (\ref{4.75}), as the formula in Eq.\ (\ref{4.1}) requires, and obtain  
\begin{eqnarray}
\sigma &\equiv& \sigma_{n\kappa \mu}=\frac{\alpha^2 Z}{N_{n\kappa}^2(4\kappa^2-1)} 
\Bigg[ 
\kappa^2-\frac{\eta_{n\kappa}^{(+)}}{4}-\frac{\eta_{n\kappa}^{(-)}}{4} + \frac{\mu^2}{4\kappa^2-1}
\nonumber \\
&& \times
\left(
\frac{2\kappa+1}{2\kappa-1}\eta_{n\kappa}^{(+)}+\frac{2\kappa-1}{2\kappa+1}\eta_{n\kappa}^{(-)} + 4\kappa^2\frac{4\kappa^2-5}{4\kappa^2-1}+ \frac{32\kappa^3[2\kappa(n+\gamma_{\kappa})-N_{n\kappa}](\alpha Z)^2}{\gamma_{\kappa}(4\gamma_{\kappa}^2-1)N_{n\kappa}^2}
\right)
\Bigg].
\label{4.78}
\end{eqnarray}
The above result determines the nuclear shielding of the relativistic hydrogenlike atom being in an \emph{arbitrary} discrete energy eigenstate, characterized by the set of quantum numbers $\{n, \kappa, \mu\}$. We are not aware of any publications, in which the analytical formula for $\sigma$ has been obtained in a such general overview. 

To test the correctness of the expression for the magnetic shielding constant derived by us in the present paper, we shall check its form for some particular atomic states and compare them with the corresponding formulas found earlier by other authors. Thus, for states of the same symmetry as the atomic ground state, i.e., with $\kappa=-1$ and $\mu=\pm1/2$, but with an arbitrary radial quantum number $n$, and --- consequently --- the principal quantum number ${N}$, Eq.\ (\ref{4.78}) takes the following form:
\begin{equation}
\sigma_{n,-1,\pm \frac{1}{2}}=\frac{2}{27} \frac{\alpha^2 Z}{N_{n,-1}^2} 
\left[
4+\frac{N_{n,-1}}{\gamma_1+n+N_{n,-1}}+\frac{12(2\gamma_1+2n+N_{n,-1})(\alpha Z)^2}{\gamma_1(4\gamma_1^2-1)N_{n,-1}^2} 
\right],
\label{4.79}
\end{equation} 
in agreement with the formula derived by Ivanov \emph{et al.} in Ref.\ \cite{Ivan09}. In turn, if one puts into our general formula $n=0$,  the expression for the magnetic shielding constant for such states is
\begin{equation}
\sigma_{0 \kappa \mu}=\frac{\alpha^2 Z}{4} 
\left[
 \frac{(2\kappa-1)^2-4\mu^2}{\kappa(2\kappa-1)^3} \: \frac{\gamma_{\kappa}-\kappa+1}{\gamma_{\kappa}-\kappa}
+\frac{(2\kappa+1)^2-4\mu^2}{\kappa(2\kappa+1)^3} \: \frac{\gamma_{\kappa}+\kappa+1}{\gamma_{\kappa}+\kappa}
+\frac{32 \mu^2}{(4\kappa^2-1)^2} \frac{4\kappa^2-2\gamma_{\kappa}^2-\gamma_{\kappa}}{\gamma_{\kappa}(2\gamma_{\kappa}-1)}
\right].
\label{4.80}
\end{equation}
The above formula is identical to the one derived by Moore in Ref.\ \cite{Moor99}, with the use of a completely different method. A very special case of the two previously described classes of atomic states is the ground state, for which both of Eqs.\ (\ref{4.79}) and (\ref{4.80}) simplify to
\begin{equation}
\sigma_{0,-1,\pm \frac{1}{2}} \equiv \sigma_{g}=-\frac{2 \alpha^2 Z}{27} \frac{4\gamma_1^3+6\gamma_1^2-7\gamma_1-12}{\gamma_1(\gamma_1+1)(2\gamma_1-1)}.
\label{4.81}
\end{equation}
This result fully conforms with earlier findings obtained by other authors in Refs.\ \cite{Zapr81, Zapr85, Zapr74, Stef12, Pype99}. Using the definition stated in Eq.\ (\ref{2.7}), it may be rewritten in the following form:
\begin{equation}
\sigma_g=\frac{2}{27 Z} \frac{4\gamma_1^4+2\gamma_1^3-13\gamma_1^2-5\gamma_1+12}{\gamma_1(2\gamma_1-1)}.
\label{4.82}
\end{equation}
Proceeding in a similar way, employing the general result given in Eq.\ (\ref{4.78}), we can find the expressions for the magnetic shielding constant for some states belonging to the first set of excited states. Thus, for the state $2p_{1/2}$ we have 
\begin{equation}
\sigma_{1,1,\pm \frac{1}{2}}=\frac{2}{27 Z} \frac{N_{1,1}(4\gamma_1^3-6\gamma_1^2+11\gamma_1-6)-2(8\gamma_1^4-18\gamma_1^3+4\gamma_1^2+9\gamma_1-6)}{\gamma_{1}(4\gamma_1^2-1)},
\label{4.83}
\end{equation}
while for the state $2p_{3/2}$ we obtain
\begin{equation}
\sigma_{0,-2,\pm \frac{1}{2}}=-\frac{2}{3375 Z} \frac{176\gamma_2^4+88\gamma_2^3-479_2^2+428\gamma_2-960}{\gamma_2(2\gamma_2-1)} \qquad  \left(\textrm{with } \ \mu=\pm \frac{1}{2}\right),
\label{4.84}
\end{equation}
\begin{equation}
\sigma_{0,-2,\pm \frac{3}{2}}=-\frac{2 \alpha^2 Z}{125} \frac{8\gamma_2^3+20\gamma_2^2-67\gamma_2-160}{\gamma_2(\gamma_2+2)(2\gamma_2-1)}
=\frac{2}{125 Z} \frac{8\gamma_2^4+4\gamma_2^3-107\gamma_2^2-26\gamma_2+320}{\gamma_2(2\gamma_2-1)}\qquad  \left(\textrm{with } \ \mu=\pm \frac{3}{2}\right).
\label{4.85}
\end{equation}
The last three equations agree with the corresponding formulas derived by Pyper and Zhang in Ref.\ \cite{Pype99}. Since the analytical expressions for the magnetic shielding constant for the given atomic states --- as well as the general formula in Eq.\ (\ref{4.78}) --- have elementary forms, we decided not to present here any tables with the numerical values of that quantity. However, during preparation of this paper, we have performed such numerical calculations for some states of the atom. The values obtained by us with the use of the formula from Eq.\ (\ref{4.78}) fully confirm with data tabulated in Refs.\ \cite{Moor99} and \cite{Pype99}; the reader may easily check the consistency of these results for any ions with the nuclear charge numbers from the range $1 \leqslant Z \leqslant 137$.

\section{Conclusions}
\label{V}
\setcounter{equation}{0}

In this work, we derived analytically a closed-form expression for the dipole magnetic shielding constant of the relativistic hydrogenlike atom in an arbitrary discrete energy eigenstate. In contrast to the results obtained by us for some other electromagnetic properties (the magnetizability \cite{Stef15} and the magnetic-dipole-to-electric-quadrupole cross susceptibility \cite{Stef16}) of Dirac one-electron atom in any excited state, the formula for the physical quantity considered here does not include any special functions, but is of an elementary form. From our general result, we obtained the expressions for the magnetic shielding constant for some particular atomic states. For states with zero radial quantum number, we reconstructed the corresponding formula derived by Moore \cite{Moor99}, from which, consequently, we received the formula for the magnetic shielding constant of the ground state of the atom, in agreement with the results predicted by several other authors \cite{Zapr81, Zapr85, Zapr74, Stef12, Pype99, Chen09, Szmy11}. Furthermore, for states of the same symmetry as the atomic ground state, i.e., for those with $\kappa=-1$, our general expression for $\sigma$ confirms the corresponding formula found by Ivanov \emph{et al.\/} \cite{Ivan09}. The results obtained by us for some states belonging to the first set of excited states, which --- in general --- do not include to the aforementioned classes of atomic states (i.e., states $2p_{1/2}$ and $2p_{3/2}$, with all possible values of the magnetic quantum number), agree with the predictions of Pyper and Zhang presented in Ref.\ \cite{Pype99}.

\begin{acknowledgments}
I am indebted to Professor R.\ Szmytkowski for many hours of valuable discussions and for commenting on this manuscript.
\end{acknowledgments}

\end{document}